\documentclass[twoside,twocolumn,leqno]{article}
\usepackage{siamproceedings}

\usepackage[leqno]{amsmath}
\usepackage{amssymb}

\newcommand{\OO}{\mathcal{O}}

\newcommand{\vecr}{\bm{r}}
\newcommand{\f}{\bm{f}}
\newcommand{\x}{\bm{x}}
\newcommand{\y}{\bm{y}}
\newcommand{\z}{\bm{z}}
\newcommand{\w}{\bm{w}}

\newcommand{\ones}{\bm{1}}
\newcommand{\matA}{\bm{A}}
\newcommand{\matC}{\bm{C}}
\newcommand{\matL}{\bm{L}}
\newcommand{\matU}{\bm{U}}
\newcommand{\matB}{\bm{B}}

\newcommand{\LI}{\mathtt{L1}}
\newcommand{\LOA}{\mathtt{L0A}}
\newcommand{\LOB}{\mathtt{L0B}}
\newcommand{\UB}{\mathtt{UB}}
\newcommand{\GM}{\mathtt{GM}}
\newcommand{\LOC}{\mathtt{L0C}}

\newcommand{\ScanU}{\texttt{ScanU}}
\newcommand{\ScanULONE}{\texttt{ScanUL1}}
\newcommand{\MCScan}{\texttt{MCScan}}
\newcommand{\MCScanULONE}{\texttt{MCScanUL1}}

\usepackage[noend]{algpseudocode}
\usepackage{bm}
\usepackage{booktabs}
\usepackage{listings}
\usepackage{color}
\usepackage[normalem]{ulem}
\usepackage{tikz}
\usepackage{caption}
\usepackage{subcaption}
\usepackage{hyperref}

\newcommand{\vnlfigwidth}{7cm}

\hypersetup{  
     unicode=false,          
     pdftoolbar=true,        
     pdfmenubar=true,        
     pdffitwindow=true,      
     pdftitle={Parallel Scan on Ascend AI Accelerators},
     pdfauthor={Anonymous}, 
     pdfsubject={Parallel Algorithms},   
     pdfcreator={Anonymous},   
     pdfproducer={Anonymous}, 
     pdfkeywords={prefix sum,scan, matrix multiplication,tensor core unit model, tensor cores, systolic arrays}, 
     pdfnewwindow=true,      
     colorlinks=true,       
     linkcolor=blue,          
     citecolor=blue,        
     filecolor=magenta,      
     urlcolor=blue,          
}

\algblock{ParFor}{EndParFor}
\algnewcommand\algorithmicparfor{\textbf{parfor}}
\algnewcommand\algorithmicpardo{\textbf{do}}
\algnewcommand\algorithmicendparfor{\textbf{end\ parfor}}
\algrenewtext{ParFor}[1]{\algorithmicparfor\ #1\ \algorithmicpardo}
\algrenewtext{EndParFor}{\algorithmicendparfor}
\algrenewtext{EndParFor}{\algorithmicendparfor}

\usepackage{amsfonts}
\usepackage{graphicx}
\usepackage{epstopdf}
\usepackage{enumitem}
\ifpdf
  \DeclareGraphicsExtensions{.eps,.pdf,.png,.jpg}
\else
  \DeclareGraphicsExtensions{.eps}
\fi

\makeatletter
\renewcommand{\ALG@name}{\sc Algorithm}
\makeatother

\usepackage{xcolor}
   
\usepackage{amsopn}

\begin{document}

\newcommand\relatedversion{}
\renewcommand\relatedversion{\thanks{The full version of the paper can be accessed at \protect\url{https://arxiv.org/abs/0000.00000}}} 

\title{\Large Parallel Scan on Ascend AI Accelerators}

\author{Bart\l omiej Wr\'{o}blewski\thanks{Work done while the author was employed at Huawei Zurich Research Center.}
\and Gioele Gottardo\footnotemark[2]    
\and Anastasios Zouzias\thanks{Computing Systems Lab, Huawei Zurich Research Center, Zurich, Switzerland
(\email{anastasios.zouzias@huawei.com}).}}


\date{}

\maketitle


\fancyfoot[R]{\scriptsize{Copyright \textcopyright\ 20XX by SIAM\\
Unauthorized reproduction of this article is prohibited}}




\begin{abstract}
We design and implement parallel prefix sum (scan) algorithms using Ascend AI accelerators. Ascend accelerators feature specialized computing units—the cube units for efficient matrix multiplication and the vector units for optimized vector operations. A key feature of the proposed scan algorithms is their extensive use of matrix multiplications and accumulations enabled by the cube unit. To showcase the effectiveness of these algorithms, we also implement and evaluate several scan-based operators commonly used in AI workloads, including sorting, tensor masking, and top-$k$ / top-$p$ sampling.
Our single-core results demonstrate substantial performance improvements, with speedups ranging from $5\times$ to $9.6\times$ compared to vector-only implementations for sufficiently large input lengths. Additionally, we present a multi-core scan algorithm that fully utilizes both the cube and vector units of Ascend, reaching up to 74.9\% of the memory bandwidth achieved by memory copy. Furthermore, our radix sort implementation, which utilizes matrix multiplications for its parallel splits, showcases the potential of matrix engines to enhance complex operations, offering up to $3.3\times$ speedup over the vector-only baseline.

\end{abstract}

\maketitle

%
\section{Introduction}\label{sec:intro}
%
%
Parallel scan is a fundamental parallel computing paradigm with many applications~\cite{blelloch_prefix_1990,book:scan94}. Due to its importance, parallel scan has been studied in several models of computation, including the circuit (work and depth) and the Parallel Random-Access Machine (PRAM) model~\cite{prog:blelloch_cacm96}. In the circuit model, the depth and size trade-offs for parallel optimal prefix circuits are well-understood for binary operations~\cite{scan:snir86,scan:zero_def_circuits06}, but other models are yet to be explored, especially in the heterogeneous computing domain~\cite{scan:tcuICS19,scan:zouzias_europar23}.
Despite many attempts, there is a large gap between abstract parallel machine models and the current state-of-the-art accelerators that contain heterogeneous computing units, see~\cite{search_pram_matias97} and 
\cite{model:PEM:arge08,provably_gpu_algos,gpu_model_ipdps14}. For example, hardware vendors have manufactured accelerators with specialized compute engines known as \emph{matrix engines} or \emph{tensor core units}. A list of such specialized hardware units includes Google's TPUs~\cite{google_tpus:ISCA17,tpu:v4}, Nvidia's Tensor Cores~\cite{nvidia:tesla_v100}, AMD's Matrix Cores~\cite{amd_matrix_cores} and Huawei's Ascend Cube Unit~\cite{huawei_ascend:hpca2021,huawei_ascend:hotchips19} to name a few. Moreover, in the CPU domain, examples of such units (or extensions) are ARM's Scalable Matrix Extension (SME)~\cite{armv9:manual_2024}, Intel's Advanced Matrix Extensions (AMX)~\cite{intel_isa_extensions} and IBM's POWER10 Matrix-Multiply Assist (MMA)~\cite{ibm_power10_mma}. Therefore, today's high-performance processors contain matrix engines that allow efficient multiplication of constant-sized matrices, making ideas from the early 1980s~\cite{kung1978systolic,algo_parallel_computers1985} a reality and initiating fruitful debates within the community~\cite{matrix_engines_debate}.
Given the presence of matrix engines in computing systems, a model of computation was recently proposed to capture matrix multiplication accelerators called the Tensor Core Unit (TCU) model~\cite{tcu_model:spaa20,tcu_model:europar21}. The authors of~\cite{tcu_model:europar21} initiated the study of TCU algorithms and revisited classical paradigms in the TCU model. The study of accelerating prefix sum (and reduction) operations using matrix multiplication units was first initiated in the seminal paper of~\cite{scan:tcuICS19}. A follow-up work presented a logarithmic-depth parallel scan algorithm in the TCU model of computation, where only matrix multiplication operations are required~\cite{scan:zouzias_europar23}. Although the main algorithm in~\cite{scan:zouzias_europar23} offers stronger theoretical guarantees than that of~\cite{scan:tcuICS19}, its poor memory access patterns limit its practical performance.
Here, we focus on simpler algorithms inspired by~\cite{scan:tcuICS19} that achieve performance close to hardware peak. Our top-performing multi-core scan kernels combine ideas from~\cite[Alg.~6]{scan:tcuICS19}, explicit software-managed asynchronous pipelining between matrix and vector cores, and a novel matrix-vector recomputation strategy in the up-sweep scan phase. Furthermore, we demonstrate that our proposed scan kernels can be easily composed to deliver the best-performing kernels for AI workloads on Ascend.
In short, we draw inspiration from both these works~\cite{tcu_model:europar21,scan:tcuICS19} to advance the study of parallel scans on matrix multiplication accelerators. As a case study and an evaluation environment, we use Huawei's Ascend AI accelerators and the Compute Architecture for Neural Networks (CANN) software ecosystem of Ascend to evaluate our proposals~\cite{book:ascend_cann_xiaoyao20}.

Our main contributions are listed below:
\begin{itemize}
    \item
    We design, implement, and evaluate parallel scan algorithms specialized for the Ascend AI accelerator. The distinguishing feature of the proposed algorithms is the extensive use of the Ascend matrix multiplication engines (``Cube cores''). In particular, we implement several scan variants, including single-core and multi-core scans, scans on multiple arrays (batched scan), inclusive/exclusive scans and specialized scan implementations for boolean (mask) inputs using the cube unit's 8-bit integer capabilities.
    \item 
    We evaluated the proposed scan algorithms that use a single cube and vector units, demonstrating a $5\times$ up to $9.6\times$ speed-up compared to the baseline scan (vector unit only).
    \item
    We present multi-core scan algorithms that fully utilize both the cube and vector units of the Ascend 910B4 accelerator, reaching up to 74.9\% of the theoretical memory bandwidth of memory copy. Moreover, the \MCScan~algorithm (\Cref{alg:scan_multi_core}) attains a $15.2\times$ speedup over the corresponding single-cube variant when utilizing all 20 available AI cores.
    \item
    We implement a list of computational kernels (operators), including parallel split, compress/compact and top-$p$ (nucleus) sampling essential to AI workloads. In all of these cases, we demonstrate significant performance improvements.
    \item
    Lastly, we implement radix sort, whose parallel splits take advantage of the cube units. The radix sort implementation provides up to $3.3\times$ speed-up over baseline sorting that does not utilize the cube units. 
\end{itemize}
%
Next, we briefly emphasize the key distinction between the above contributions and prior art. First, the multi-core scan algorithm is novel (to the best of our knowledge) since it performs partial recomputation of the reduction values on both the cube and vector units in its first phase. This recomputation strategy is different compared to all previously known scan strategies on accelerators, see Section~\ref{sec:background_scan}. Such a recomputation strategy could be of interest to other matrix multiplication accelerators as well.
Second, our radix sort implementation yields an intriguing result: in practice, multiple small dense matrix multiplications can be leveraged to improve the end-to-end performance of parallel sorting. Although the algorithmic ideas and techniques underlying this result are well-established~\cite{radix_sort:spaa91_CM2}, we believe that it paves the way for interesting research directions in the future. As an example, we pose the following question, see also the presentation of~\cite{tcu_model:spaa20}. Is it possible to also utilize the multiple-add capabilities of the matrix multiplications units to improve further the performance of parallel sorting? It is important to note that, despite relying on the matrix multiplications for scan operations during parallel splits, our radix sort does not exploit the multiply-add capability of the matrix unit.
%
\section{Background}\label{sec:background}
%
In this section, we give an overview of related work in computing parallel scans using accelerators, focusing mostly on matrix multiplication accelerators. We also present the Tensor Core Unit (TCU) model of computation, and we give a summary of a selected list of scan-based primitives. The inclusive prefix sum of a sequence of elements $x_1,x_2,x_3,\dots$ equipped with a binary associative operator $\odot$ is the sequence $x_1, x_1\odot x_2, x_1\odot x_2\odot x_3, \dots$.
%
\subsection{Scan Strategies on Accelerators}\label{sec:background_scan}
%
Numerous parallel scan implementations and libraries have been proposed in the literature~\cite{scan:matrixscan08,streamscan_ppopp13,book:gpus:scan_chapter, scan:gpu_harris07,scan:moderngpu,cuda:thrust,nvidia:cub,scan:decouple_lookback}. Here, we discuss the most relevant work focusing on scan implementations targeting accelerators, most notably Graphics Processing Units (GPUs). Horn was one of the first to implement parallel scans in GPUs~\cite[Chapter~36]{gpu_gems_book_2}, followed up with several improvements.
In a nutshell, GPU scan implementations primarily follow a two-level hierarchical approach where the highest level of the hierarchy is the block level. An efficient scan implementation follows one of these scan strategies: \emph{Scan-Scan-Add (SSA)}, \emph{Reduce-Scan-Scan (RSS)},
or \emph{Stream-scan} according to the state-of-the-art \emph{decouple lookback scan} approach of~\cite{scan:decouple_lookback}. Scan-Scan-Add (SSA) means that the (local) \emph{scans} are initially computed per block. Second, the values of the largest index of each block are collected and \emph{scanned}. Third, the collected per-block scan values are broadcast-\emph{added} to their corresponding blocks.  Similarly, the RSS approach makes a block-level reduction first, followed by a scan of the block-level reductions and a final scan of each block. We refer the interested reader to~\cite[Section~3]{scan:decouple_lookback} for a more detailed discussion.
Scan is a memory-bound operation and, hence, the main drawback of the SSA and RSS scan strategies is the high number of elements that are read and written to global memory. In particular, for input length $N$, SSA reads/writes $\approx 4N$ elements, whereas RSS reads/writes $\approx 3N$ elements. Such a reduction in the memory access size is critical for improving the performance of scans.

StreamScan and decouple look-back strategies access only $~2N$ memory elements but need to efficiently handle the sequential data dependency of the scan computation using adjacent (block) synchronization, see~
\cite[Chapter~11.7]{book:gpus}. StreamScan is a single-pass approach in which each thread block is assigned a tile of input, and a serial dependency between the blocks exists~\cite{streamscan_ppopp13}. The critical feature of StreamScan is that it requires synchronization between adjacent blocks only (without global block-level synchronization). The decoupled look-back strategy of~\cite{scan:decouple_lookback} aims to alleviate the drawbacks of the serial dependency of StreamScan by performing redundant work to ``dissociate'' local computation from the latency of global prefix propagation. Both StreamScan and decouple look-back strategies require only $~2N$ data movement to global memory: $N$ input elements are read, $N$ output elements are written.
To the best of our knowledge, the state-of-the-art scan algorithm on GPUs follows a highly-tuned decoupled look-back and backoff approach, which is briefly explained in~\cite{NaN2024ScanSpeed}.
%
%
%

%
\subsection{Tensor Core Unit (TCU) Model}\label{sec:background:models}
%
To the best of our knowledge, the Tensor Core Unit (TCU) model is the only model of computation that has been recently proposed to capture matrix multiplication accelerators~\cite{tcu_model:spaa20}. The TCU model is a standard RAM model with an additional circuit, named tensor core unit, that performs matrix multiplication between constant-size matrices. Although the TCU model captures well a single matrix multiplication computational unit of today's accelerators, it ignores other essential features: the presence of vector processing units and, more critically, the multi-core nature of these accelerators. Since the TCU model considers only a single matrix multiplication unit, it does not allow its users to conduct a work/depth algorithmic analysis~\cite{prog:blelloch_cacm96}. Due to the above limitations, any algorithmic analysis in the TCU model will not correspond to a realistic execution in Ascend, i.e., ignoring parallelism and the vector units. Nevertheless, we discuss the work/depth asymptotic analysis of the proposed algorithms, assuming the presence of multiple matrix engines and vector units, considering their operations as basic operations.
%
\subsection{Applications of Parallel Scan}\label{sec:background_scan_apps}
%
Parallel scan has a plethora of applications~\cite{book:scan94,blelloch_prefix_1990}.
Here, we restrict our attention to scan applications that enable us to generate efficient computational kernels (operators) that appear in AI workloads. In particular, we have identified that sorting, weighted sampling, masking of tensors, and top-$k$/top-$p$ sampling are essential. All these applications of the parallel scan are well-known, but the observation that top-$p$ sampling can benefit from scan seems to be new (to the best of our knowledge).
%
\section{Ascend AI Accelerators}\label{sec:matmul_acc}
%
In this section, we briefly discuss the DaVinci architecture of Ascend accelerators consisting of the cube and vector computing units, mostly following~\cite{huawei_ascend:hotchips19}. Moreover, we discuss the AscendC programming model of Ascend, a recently proposed programming model for Ascend operator development. All the material presented here is available online at~\href{https://www.hiascend.com/document/detail/zh/CANNCommunityEdition/80RC3alpha002/quickstart/quickstart/quickstart_18_0001.html}{https://www.hiascend.com}.
%
\subsection{Ascend Hardware}\label{sec:ascend}
%
Huawei Ascend 910B is a recent series of Huawei chips designed to accelerate neural network training and inference. For the scope of the paper, an accelerator can be seen as a grid of computing units called AI Cores and a global High Bandwidth Memory (HBM) with L2 cache.

In the Ascend 910B series, an AI Core consists of one AI Cube (AIC) core and multiple, usually two, AI Vector (AIV) cores. Each AIC and AIV core contains a scalar unit for basic arithmetic operations, program flow control, calculating addresses, and dispatching instructions. Each core also includes computing engines (either vector or cube ones), local memory buffers, and Memory Transfer Engines (MTEs). MTEs are responsible for moving data between global and local memory buffers. Both MTEs and computing engines have separate instruction queues and work in parallel, so it is the programmer's responsibility to ensure synchronization.

An AI Vector core performs vector operations similar to traditional SIMD operations. The input and output data of an AI Vector core must be allocated to the local scratchpad/buffer called Unified Buffer (UB). AIV cores support simple arithmetic operations such as vector addition and more complex ones such as gather and reduce.

\begin{figure}[t]
\begin{center}
  \includegraphics[width=\columnwidth, scale=0.8]{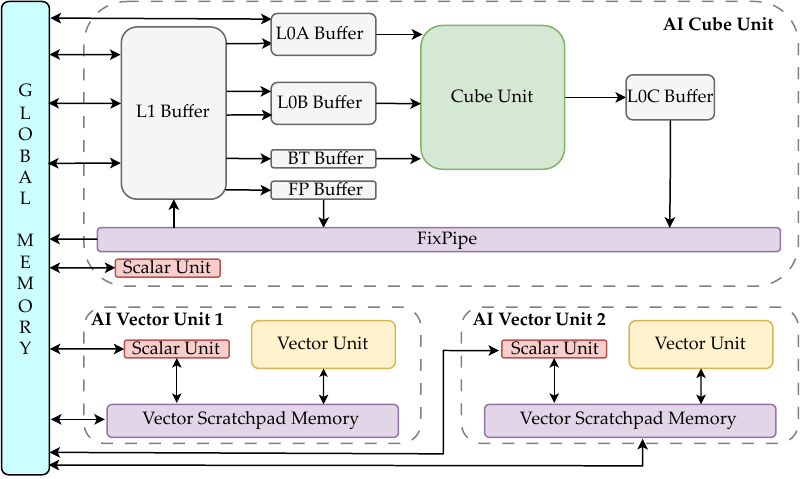}
\end{center}
\vspace{-7mm}
\caption{\small Architecture of Ascend 910B accelerators. Each AI core contains one cube and two vector units.}
\label{fig:ascend910B}
\end{figure}

An AI Cube core is primarily responsible for matrix multiplication operations. The AI Cube core contains a hierarchical scratchpad memory structure (L1, L0A, L0B, L0C, BT, FP buffers) and a cube computing engine. An AIC core can be configured to multiply two matrices of almost arbitrary sizes. It also supports result accumulation, selected activation functions, and quantization operations. The cube core supports both floating point and low-precision integer data types, i.e. float16 (with float32 output) and int8 (with int32 output).
\Cref{fig:ascend910B} shows the Ascend architecture where the Cube and Vector units are separate cores. In the 910B architecture, data can only be exchanged using global memory and/or L2 cache.
%
%
%

%
\section{Scan Algorithms on Ascend}\label{sec:algos}
%
In this section, we discuss the design and implementation of parallel scan algorithms using matrix multiplication accelerators. Although the Ascend AI accelerator is used as a case study for the implementations, we aim to decouple the fundamental algorithmic ideas from the intricate architectural details of the Ascend accelerator as much as possible.
Matrix multiplication is an essential operator in our discussions here; hence, we introduce some linear algebraic notation that will be used throughout the paper. We denote matrices using boldface font and capitals, i.e., $\matA, \matB, \matC$. We use $s$ to denote the size of square matrices, i.e., $\matA_s$, and define $\ell:=s^2$. We drop the subscript on the matrices when the dimension is clear from the context. We denote matrix multiplication between $\matA$ and $\matB$ by $\matC := \matA\ @\ \matB$. We denote by $\matU_s$ the upper-triangular all-ones square matrix of size $s$, including ones on the main diagonal. $\matL_{s}$ is the lower triangular all-ones of size $s$. $\matL^{-}_{s}$ is the \emph{strictly} lower triangular all-ones of size $s$: $\matL_{s}^{-}$ has zeroes on the main diagonal. $\ones_s$ denotes the all-ones square matrix of size $s$. We denote one-dimensional arrays using boldface font and lower-case: $\vecr,\x,\y,\z$.  We frequently partition an array $\x$ into tiles of length $\ell$, i.e., tiles are contiguous blocks of $\ell$ entries of $\x$. We note an arbitrary $\ell$-tile of $\x$ by $\x_\ell$. We also view a tile $\x_\ell$ as a row-major matrix $\matA$ having $s$ rows and $s$ columns (zero-pad if necessary).
In Section~\ref{sec:single_cube_core_scan}, we present two scan algorithms (\Cref{alg:scan_single_core} and~\Cref{alg:scan_cube_only}) that use a single matrix multiplication unit; we call these algorithms respectively \ScanU~and \ScanULONE, based on which constant matrices they use.
The key ingredient here is to utilize Ascend's cube unit effectively. In particular,~\Cref{alg:scan_cube_only} performs multiple matrix multiplications and utilizes the accumulation buffer of the Cube unit to compute the scan of an input tile of length $\ell$, whereas~\Cref{alg:scan_single_core} computes $s$ consecutive scans of smaller tiles of length $s$ using a single matrix multiplication.
In Section~\ref{sec:multi_core_scan}, we present a multi-core scan algorithm (\MCScan,~\Cref{alg:scan_multi_core}) tailored for the Ascend AI accelerator. A key feature of \MCScan~is that it utilizes all the available cube and vector cores. \MCScan~and its extensions are designed for scenarios involving very large one-dimensional arrays.
In Appendix~\ref{sec:batched_scan}, we build on top of the \ScanU~and \ScanULONE~algorithms and extend them into batched scan variants that operate on multi-dimensional arrays (tensors). We highlight that such extensions are non-trivial, as several issues arise when scheduling many scan operations across multiple cores, including padding and improving load balancing on vector/cube units with a 2:1 ratio, among others.
%
\subsection{Warm-up: single cube scans}\label{sec:single_cube_core_scan}
%
In this section, we present two scan algorithms that utilize a single cube and vector units. Both algorithms are tailored to the DaVinci architecture and are based on the linear algebra fact that if $\matA$ is the row-major matrix view with $s$ columns of a vector $\x$ then:
\begin{quote}
Matrix multiplication $\matA \ @\ \matU_s$ computes ``local'' scans of tiles of size $s$ of $\x$.
\end{quote}
%
\begin{figure}[ht]
\begin{center}
  \includegraphics[width=6cm]{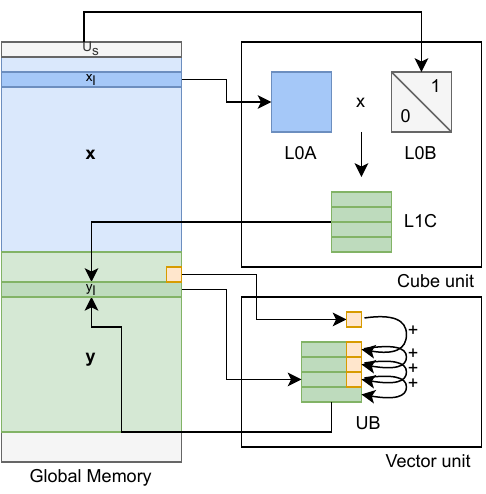}
\end{center}
\vspace{-7mm}
\caption{\small Data path from an input tile $\x_\ell$ to an output tile $\y_\ell$ of the \ScanU~(\Cref{alg:scan_single_core}). 
}
\label{fig:scan_singlecore}
\end{figure}
%
%

%
Both algorithms process their tiles using a pipelined execution strategy. The critical path or span of both proposed algorithms is linear on the input length for constant values of $s$, since there is a sequential dependency on the partial sums. Therefore, these kernels are more effective when the input array to be scanned has a relatively short length. Additionally, designing and implementing scan algorithms that use a single cube core is a building block for extending these ideas to a multi-core scenario, as seen in Sections~\ref{sec:batched_scan} and~\ref{sec:multi_core_scan}.
The first algorithm, \ScanU~(\Cref{alg:scan_single_core}), computes $s$ consecutive local scans of tiles of size $s$ using the cube unit and then propagates the partial sums using a single vector unit. More precisely, once the cube unit has computed the local row scans of a matrix tile of size $\ell$, the tile is sent to a vector core for further processing. The vector core will add a scalar to each row to correct the prefix sum. It is important to note that to obtain the correct prefix sum, the vector core keeps track of the last value of each $\ell$-tile and propagates it to the next tile.
\Cref{fig:scan_singlecore} depicts the data movements and the memory view of \ScanU. The input vector $\x$ is shown in blue. A tile $\x_\ell$ of $\x$ and $\matU_s$ are loaded into the cube unit where the matrix multiplication occurs. The matrix multiplication result is written to global memory. The vector unit reads the cube unit output tile and propagates the prefix sum in place. The whole process, which consists of memory transfers and cube/vector operations, is pipelined over the input tiles using AscendC software pipeline capability.
%
\begin{algorithm}[h]
\begin{algorithmic}[1]
\Procedure{ScanU}{$\x$, $s$}
\State Let $\y$ be the output array
\State $\textit{partial} \gets 0$ \Comment{Accumulation value (Vector Unit)}
\State Load $\matU_s$ in $\LOB$
\For {each $s^2$-tile of $\x$: $\x_\ell$} \Comment{Pipelined exec.}
\State Load $\x_\ell$ to $\LOA$ \Comment{Cube unit}
\State $\matC \gets \matA_s\ @\ \matU_s $ \Comment{acc. \textbf{off}, free $\LOA$}
\State Copy $\matC$ from $\LOC$ to $\y_\ell$ in $\GM$ 
\State Vector unit waits tile from Cube unit
\State Copy $\y_\ell$ from $\GM$ to $\UB$\Comment{Vector unit}
\For {each $s$-tile of $\y_\ell$: $\y_s$} 
\State $\y_s \gets \y_s + \textit{partial}$
\State $\textit{partial} \gets$ last entry of $\y_s$
\EndFor
\State Copy $\y_\ell$ from $\UB$ to $\GM$
\EndFor
\State Return $\y$
\EndProcedure
\end{algorithmic}
\caption{\small Scan Cube-Vector (\ScanU)}\label{alg:scan_single_core}
\end{algorithm}
%

%
The second algorithm, \ScanULONE~(\Cref{alg:scan_cube_only}), is an Ascend adaptation of~\cite[Algorithm~6]{scan:tcuICS19} and is based on a matrix identity that expresses the scan of an array $\z$ of length $\ell$ using matrix operations. View $\z$ as a square row-major matrix $\matA$ of size $s=\lceil \sqrt{\ell}\rceil$ (pad with zeroes if needed). Given $\z$, the inclusive scan of $\z$ ($\texttt{scan}(\z)$) can be computed as:
%
\begin{equation}\label{eqn:scan}
\texttt{scan}(\z) = \matA_s\ @\ \matU_s + \matL^{-}_s\ @\ \matA_s\ @ \ \ones_s,
\end{equation}
%
ignoring any padded values. Equation~\ref{eqn:scan} first appeared in~\cite{scan:tcuICS19}. \ScanULONE~uses Equation~\ref{eqn:scan} to scan each consecutive tile of size $\ell$ (Lines 6-12 of~\Cref{alg:scan_cube_only}), and then propagates the last value of the partial sums sequentially. In a high-level, for each tile of size $\ell$, the cube unit evaluates Equation~\ref{eqn:scan} with the following sequence of matrix operations:
%
{\renewcommand{\arraystretch}{0.5} 
\begin{align}
    \matC_1 &= \matA_s\ @\ \ones_s \nonumber \\
    \matC_2 &= \matA_s\ @\ \matU_s \nonumber \\
    \matC_2 &= \matC_2 + \matL_s^{-}\ @\ \matC_1. \nonumber
\end{align}
}
%
The above sequence of matrix operations has two advantageous properties with respect to data movements. The first two steps of the above sequence share the left matrix operand $\matA$, allowing us to load $\matA$ only once in $\LOA$. Moreover, the third step effectively utilizes the accumulation buffer of the cube unit since $\matC_2$ is reused in the last two steps. Once the local scan of a tile of size $\ell$ is computed, a single vector core adds the last value of the previous scanned tile to the current tile (see Lines $14-16$ of~\Cref{alg:scan_cube_only}).
%
%
\begin{algorithm}[t]
\begin{algorithmic}[1]
\Procedure{ScanUL1}{$\x$, $s$}
\State Let $\y$ be the output array
\State $\textit{partial} \gets 0$ \Comment{Accumulation value (Vector Unit)}
\State  Load $\matU_s,\matL_s^{-}, \ones_s$  in $\LI$
\For {each $s^2$-tile of $\x$: $\x_\ell$}  \Comment{Pipelined exec.}
\State Load $\x_\ell$ to $\LOA$ and $\ones_s$ to $\LOB$ \Comment{Cube unit}
\State $\matC_1 \gets \matA_s\ @\ \ones_s$ 
\State Copy $\matC_1$ from $\LOC$ to $\LI$ 
\State Load $\matU_s$ to $\LOB$
\State $\matC_2 \gets \matA_s\ @\ \matU_s$ 
\State Load $\matL_s^{-}$ in $\LOA$ and $\matC_1$ in $\LOB$
\State $\matC_2 \gets \matC_2 + \matL_s^{-}\ @\ \matC_1$ 
\State Copy $\matC_2$ from $\LOC$ to $\y_\ell$ in $\GM$ 
\State Vector unit waits tile from Cube unit
\State Copy $\y_\ell$ from $\GM$ to $\UB$ \Comment{Vector unit}
\State $\y_\ell \gets $ $\y_\ell + \textit{partial}$
\State $\textit{partial} \gets $ last entry of $\y_\ell$
\State Copy $\y_\ell$ from $\UB$ to $\GM$ 
\EndFor
\State Return $\y$
\EndProcedure
\end{algorithmic}
\caption{\ScanULONE~is adaptation of~\cite[Alg.~6]{scan:tcuICS19}}\label{alg:scan_cube_only}
\end{algorithm}
%

\begin{figure}[h]
\begin{center}
  \includegraphics[width=\columnwidth]{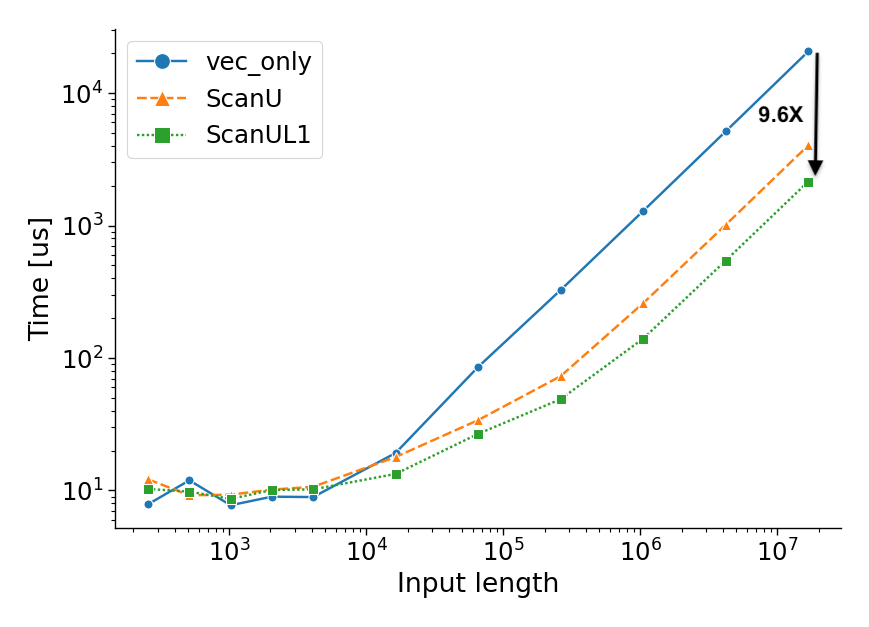}
\end{center}
\vspace{-8mm}
\caption{\small Execution time of \texttt{AscendC::CumSum} (vec\_only) versus \ScanU~and \ScanULONE~(log-log scale).}
\label{fig:vec_only_vs_cube_scan}
\end{figure}

%
\paragraph{Does cube utilization imply performance?}
%
Although the above scan algorithms demonstrate that it is possible to utilize the cube units for scan, it is unclear if cube utilization translates to performance improvements compared to ``vector only'' scan algorithms. We provide an experimental evaluation demonstrating the benefits of using the Cube unit. We developed a vector-only kernel that uses the \texttt{CumSum} AscendC API\footnote{CumSum API documentation available at \href{https://www.hiascend.com/document/detail/zh/CANNCommunityEdition/80RC3alpha003/apiref/opdevgapi/atlasascendc_api_07_0524.html}{https://www.hiascend.com} (accessed on $25$ August $2024$).} with \texttt{CumSumInfo} parameters set to $128$ and $128$. We also set $s=128$ on the cube scan algorithms to ensure a fair comparison. \Cref{fig:vec_only_vs_cube_scan} compares the cube scan algorithms and the vector-only algorithm provided by AscendC. The figure demonstrates a significant performance improvement (5$\times$ for \ScanU, and 9.6$\times$ for \ScanULONE) compared to the vector-only \texttt{CumSum} algorithm. Moreover, the figure shows that the \ScanULONE~scan algorithm has roughly a $2\times$ speedup compared to \ScanU. The critical insight here is that a more sophisticated usage of the computational capabilities of the cube unit can deliver further significant performance improvements.
%
%
%
\paragraph{Work/Span Analysis of Single Core Scans.}
%
We bound the work and span of the single-core algorithms. For work, we separately report the number of square matrix multiplications of size $s$ and the sum of lengths of the vector instructions similar to the Vector-RAM model~\cite{book:blelloch_vector_models}. The number of matrix multiplications/accumulations for \ScanU~ and \ScanULONE~are $\lceil n/s^2 \rceil$ and $3\lceil n/s^2 \rceil$, respectively. The total vector work for both algorithms is $\OO(n)$.
Next, we bound the span of the algorithms. The number of vector instructions called by the algorithms \ScanU~and \ScanULONE~are $\OO(n/s)$ and $\OO(n/s^2)$, respectively. Notice that the vector length of the \ScanULONE~vector operations is $\OO(s^2)$. Since the vector instructions are on the critical path, the span of \ScanU~and \ScanULONE~is $\OO(n/s)$ and $\OO(n/s^2)$, respectively. In summary, \ScanULONE~has a $\OO(s)$-depth speed-up compared to \ScanU, but, in practice, the achievable speed-up is $\approx 1.92$ for $s=128$, see~\Cref{fig:vec_only_vs_cube_scan}.
%
\subsection{Multi-core Scan}\label{sec:multi_core_scan}
%
This section presents a multi-core scan algorithm (\MCScan,~\Cref{alg:scan_multi_core}) to compute the prefix sum of large input arrays. For the sake of clarity, the presented algorithm employs a 1-to-1 vector-to-cube core ratio. Our implementation takes advantage of the 2-to-1 ratio of 910B, but we consider it an implementation detail.
\MCScan~is similar to the Scan-Scan-Add (SSA) paradigm of computing scans using a hierarchical partition of the input into blocks (top-level of the hierarchy) and tiles within each block, as discussed in Section~\ref{sec:background_scan}. \MCScan~consists of two phases separated by a global synchronization barrier across all cores (blocks). In contrast to previous methods, \MCScan~performs partial re-computation of the block-level reductions during its first phase, as we will explain next.
In the first phase, the cube and vector units work in parallel to partially compute the first scan part of SSA simultaneously. The cube units compute the local prefix sums of all consecutive (row) tiles of size $s$ and write them back to global memory. In parallel with the cube units, the vector units compute the reduction over the tiles and then hierarchically reduce the tile reductions on a block-level granularity. The result of these reductions is written in an array $\vecr$ where the $i$-th entry equals the reduction of all the values of the $i$-th block. By definition, $\vecr$ has length equal to the number of blocks, $B$.
In the second phase, the vector units read the local $s$-tile scans and reduction values per block $\vecr$ from global memory. First, every vector core independently computes the prefix sum of the array $\vecr$ in its local scratchpad memory $\UB$, i.e., performs a ``small'' scan on the block-level reduction. Next, each vector core uses the scanned reduction values to propagate (add) the results of the local $s$-tiled scans.
%
%
\begin{algorithm}[h!]
\begin{algorithmic}[1]
\Procedure{\MCScan}{$\x$, $s$, $B$} \Comment{$B$: number of blocks}
\State Let $\y$ be the output array
\State Let $\vecr$ be an array of length $B$ in $\GM$
\ParFor {$i$-th block of $\x$: $\x[i]$}\Comment{Phase I}
\State Load $\matU_s$ in $\LOB$ \Comment{Cube Units}
\For {each $\ell$-tile of $\x[i]$: $\x_\ell$}
\State Load $\x_\ell$ from $\GM$ to $\LOA$ 
\State $\matC \gets \matA_s\ @\ \matU_s$ 
\State  Copy $\matC$ in $\y[i]$ in $\GM$ 
\EndFor
\State Load $\x[i]$ to $\UB$ \Comment{Vector Units}
\State $r_i\gets $ \Call{ReduceSum}{$\x[i]$} 
\State Write $r_i$ on $i$-th entry of $\vecr$ in $\GM$
\EndParFor
\State SyncAll: Synchronize all cube/vector cores
\ParFor {$i$-th block of $\y$: $\y[i]$} \Comment{Phase II}
\State Load $\vecr$ from $\GM$ to $\UB$
\State $\textit{partial} \gets$ Sum first $i$ entries of $\vecr$
\For {each $\ell$-tile of $\y[i]$: $\y_{\ell}$}
\For {each $s$-tile of $\y_{\ell}$: $\y_s$}
\State $\y_s \gets \y_s + \textit{partial}$
\State $\textit{partial} \gets$ last entry of $\y_s$
\EndFor
\State Copy $\y_\ell$ from $\UB$ to $\GM$
\EndFor
\EndParFor
\State Return $\y$
\EndProcedure
\end{algorithmic}
\caption{Multi-core Scan (\MCScan)}\label{alg:scan_multi_core}
\end{algorithm}
%
%
\paragraph{Work/Span Analysis of \MCScan.}
%
First, we bound the work of \MCScan~in terms of the number of matrix multiplications of size $s$, and the number of vector instructions (scalar-vector addition) of length $s$. The number of matrix multiplications of \MCScan~is $\lceil n/s^2\rceil$. The vector instructions of length $s$ consist of $\OO(n/s)$ reductions in the first phase, and $\OO (n/s)$ scalar-vector additions in the second phase. Hence, the total number of vector operations is $\OO(n/s)$. The critical path of \MCScan~is on the second phase where the vector units propagate the partial sums, thus the span of \MCScan~is $\OO(\frac{n}{sB})$.
%
\paragraph{Exclusive scan and int8 support.}
%
We have added support for exclusive scans and input with int8 data type to our \MCScan. The interested reader is referred to Appendix~\ref{appendix:exclusive_mcscan_int8} for more details.
%
\section{Operators based on Scan}\label{sec:ops}
%
In this section, we revisit several computational parallel primitives based on parallel scans~\cite{scan:blelloch_iee89,scan:gpu_harris07}. Scan-based primitive include weighted sampling, split and compress/compact (equivalent to the \texttt{masked\_select} Pytorch operator). It is well-known that radix sort can be implemented on top of split~\cite[Section~1.3]{blelloch_prefix_1990}. Also top-k can be implemented on top of split using a partial quick-sort/select approach~\cite{prog:blelloch_cacm96}.
%

%
Interestingly enough, current AI workloads like Large Language Model (LLM) inference make implicitly heavy use of scan-based computational primitives, including top-$k$ and top-$p$ (nucleus) sampling~\cite{top_p_sampling}, see also~\cite{gu2024mamba}. The top-$p$ sampling implementation of the popular open-source model Llama3~\cite{llama} contains a batched sorting and prefix sum operation as the first two PyTorch operations, see~\cite{llama_generation_2024}.
%
\begin{figure}[ht]
\begin{center}
\includegraphics[width=0.55\columnwidth,scale=0.55]{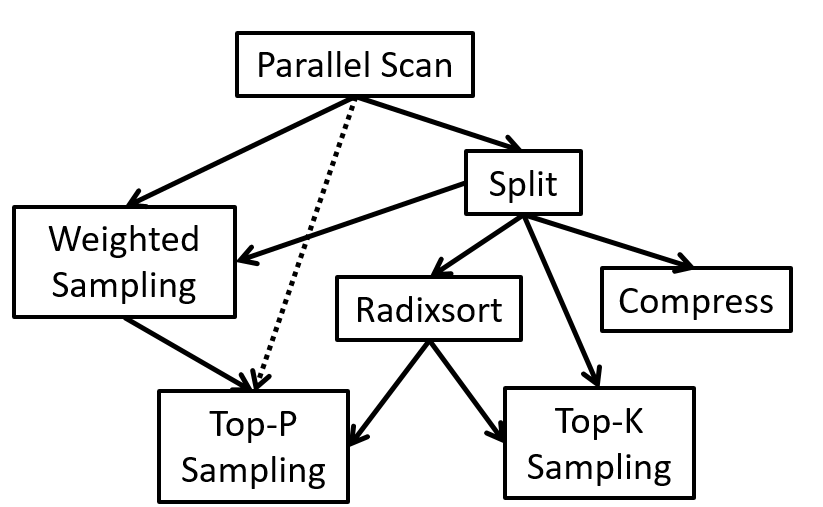}
\end{center}
\vspace{-7mm}
\caption{\small A diagram of well-known parallel scan applications considered here along with their dependencies.}
\label{fig:scan_apps_deps}
\end{figure}
%

%
In the rest of the section, we describe the split, compress, radix sort, top-$k$, top-$p$ sampling and weighted sampling operators in more detail.~\Cref{fig:scan_apps_deps} depicts the dependencies between the scan-based primitives. Top-$p$ sampling extensively uses multiple scan operations, as discussed in Section~\ref{sec:apps:top_p}. 
%
\subsection{Split}
%
The split operation takes as input an array $\x$ and a boolean flag array $\f$ of the same length. Split reorganizes the elements of $\x$ into an output array $\z$ as follows. It places all items of $\x$ where the corresponding flag is true at the beginning of the output array, followed by all items where the corresponding flag is false. One crucial property of split is the relative order of the elements is preserved, i.e., \emph{stable} ordering.
We implemented a split AscendC operator \texttt{SplitInd} that also returns the output indices corresponding to the original input locations. The output indices of \texttt{SplitInd} allow us to implement a sorting algorithm that satisfies the PyTorch API of \texttt{sort()}, which also returns the indices. \texttt{SplitInd} takes as input an array of 16-bit elements and a 0/1 mask array (flags are stored in int8). \texttt{SplitInd} executes an exclusive scan using \MCScan~on the mask array. Afterwards, it gathers the correct input elements and their indices, using vector core's \texttt{GatherMask} instruction and it stores them in global memory at the offsets calculated by the scan.
%
\subsection{Compress}
%
Compress is a particular case of split in which only the first part of the output elements of the split are returned. We have implemented a compress kernel that internally uses the exclusive \MCScan~algorithm on the mask array whose data type is $8$-bit integers. Compress is equivalent to the PyTorch \texttt{torch.masked\_select} operator that we use as a baseline for comparison in the experimental section.
%
\subsection{Radix Sort}
%
Radix sort is a well-known application of the split operator~\cite{blelloch_prefix_1990,radix_sort:spaa91_CM2}. A radix sort algorithm loops over the bits of the input elements, starting at the least significant bit and executes a split where the mask is obtained by reading the corresponding bit (radix) on each iteration. We implement a Least-Significant Bit (LSB) radix sort in AscendC using the split operator based on the \MCScan~algorithm. We implemented an additional vector-only kernel, \texttt{RadixSingle}, that extracts the radices of the inputs before the execution of the split. \texttt{RadixSingle} makes use of the AscendC vector instructions \texttt{ShiftRight} and \texttt{Not} to create the input mask for split. Additional pre-processing and post-processing phases are needed to support floats; see Exercises 8 and 9 in~\cite[Section~5.2.5]{book:knuth:artv3_sorting}. The pre-processing phase encodes all the input elements by inverting the Most Significant Bit (MSB) of positive numbers and all the bits of the negative numbers. Applying an unsigned integer radix sort on the encoded elements will correctly order them. The post-processing phase is needed to decode the elements back to the original value. We have implemented the pre- and post-processing steps using AscendC bit-wise vector instructions, and thus, support sorting of fp16.
The paper on evaluating radix sort using the Connection Machine (CM-2) came to our attention in the later stages of our radix sort development~\cite{radix_sort:spaa91_CM2}. In this work, the authors share several important implementation details for radix sort and floats that are replicated here.



%
\subsection{Top-$k$}
%
Top-$k$ selection is an essential operation in various settings, including similarity search queries~\cite{topK:database21} and Large Language Models (LLMs) inference where the output tokens are typically sampled from the $k$ tokens having the highest probability for the given context~\cite{topK:ICML19}. The interested reader is referred to a recent survey on parallel top-$k$~\cite{topK:survey_SC23}. A recent work on top-$k$ is Radik, a Radix-based GPU implementation that scales well for large values of $k$~\cite{topK:poster_PPoPP24,topK:radik_ICS24}.
We implemented a top-$k$ kernel using the selection (partial quicksort) algorithm based on our \texttt{SplitInd} operator and compared it against the baseline top-$k$ operator. Unfortunately, although improving the performance of the top-$k$ operator was a primary motivation for this work, we could not improve the performance of the baseline top-$k$ for small values of $k$ ($k\leq 4096$).
%
\subsection{Top-$p$ or Nucleus Sampling}\label{sec:apps:top_p}
%
Top-$p$ sampling in Large Language Model inference is an additional operation that applies sort and scan on the token probability vector~\cite{top_p_sampling}. These operations are usually batched with a constant batch size. Interestingly, if the sorting step is implemented using radix sort, the top-$p$ sampling operator becomes a scan-intensive operator! Indeed, top-$p$ executes 17 scans for each batch: 16 scan operations for radix sort (one scan per bit, fp16) and an additional scan required by the algorithm.
Due to space constraints, the study of the weighted sampling operator is deferred to the Appendix~\ref{sec:weighted_sampling}.
%
\section{Experimental Evaluation}\label{sec:experiments}
%
In this section, we evaluate the performance of a selective list of parallel scan algorithms and applications presented in the previous sections using Ascend AI Accelerators.

We evaluate the multi-core scan, compress, radix sort, batch scan, and top-$p$ sampling. We have implemented all proposed algorithms in C++17 using the AscendC programming framework discussed in Section~\ref{sec:ascendc}. We used the Ascend CANN toolkit 8.0.RC3.alpha002 with Ascend firmware and drivers versions 1.0 and 23.0.0, respectively. All evaluations are performed on Huawei's Ascend 910B4 accelerator. In particular, 910B4 contains 20 Cube Units and 40 Vector Units (the vector-to-cube units ratio is 2-to-1). The theoretical memory bandwidth of 910B4 is $800$ GB/s. The host CPU is an AMD EPYC processor running on Ubuntu 22.04.
All timing measurements are collected using the PyTorch profiler functionality in Python. We used Ascend's PyTorch adapter\footnote{\href{https://gitee.com/ascend/pytorch}{https://gitee.com/ascend/pytorch}} with version v2.1.0 to report all our PyTorch-related measurements. To wrap our custom AscendC operators in PyTorch, we used the open-sourced operator plugin framework at \url{https://gitee.com/ascend/op-plugin}. \texttt{op-plugin} allows its users to easily define a custom PyTorch operator. Before wrapping our operators using the \texttt{op-plugin}, we used the \texttt{msopgen}\footnote{\url{https://www.hiascend.com/en/document}} tool to wrap the AscendC operators.
%
\paragraph{Evaluation against other accelerators/architectures}
%
In this work, our primary motivation is to improve the performance of parallel scans and their applications on Ascend. In particular, our main objective is to design and implement algorithms that saturate Ascend's memory bandwidth. Comparing our results with architectures like GPUs or TPUs would likely translate to comparing the hardware specifications, i.e. memory bandwidth, rather than actually evaluating our algorithms. In addition, a comparison between accelerators that are not manufactured using the same technology node could lead the reader to wrong conclusions. Nevertheless, we present all our results in terms of bandwidth (GB/s or GElems/s), which allows for easy comparison with other architectures for the interested reader.
%
\subsection{Multi-core Scan (\MCScan)}
%
%
\begin{figure}[h]
\begin{center}
  \includegraphics[width=\columnwidth]{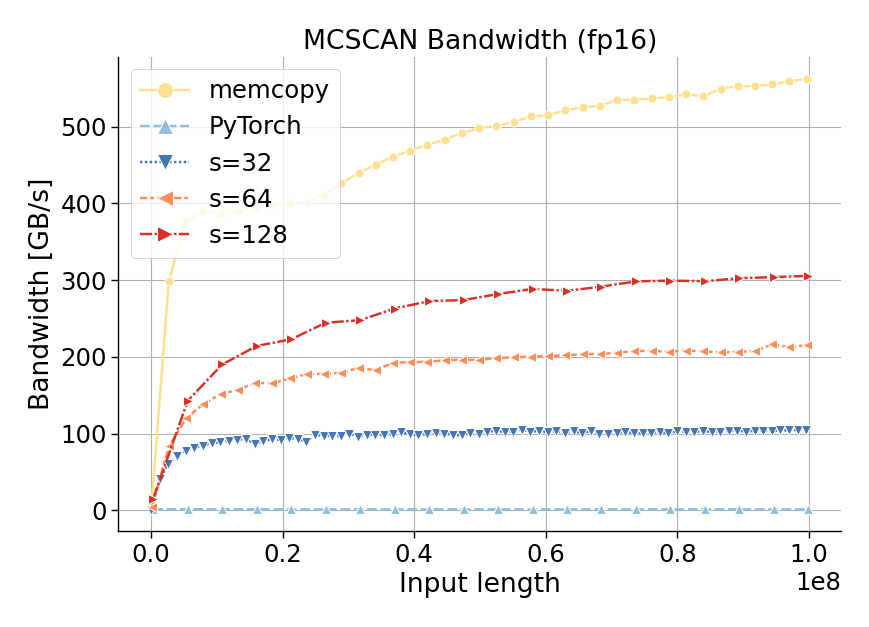}
\end{center}
\vspace{-7mm}
\caption{\small Bandwidth of \MCScan~(\Cref{alg:scan_multi_core}) for $s=32,64,128$. \MCScan~has $15.2\times$ speedup against \ScanU~on 910B4 (20 AI cores).}
\label{fig:exp:mc_scan}
\end{figure}
%
\Cref{fig:exp:mc_scan} depicts the performance of \MCScan~(\Cref{alg:scan_multi_core}) on Ascend 910B4 versus the state-of-the-art (\texttt{torch.cumsum}). We integrated the multi-core algorithm in PyTorch to ensure a fair comparison and expose it as a custom PyTorch operator. The custom PyTorch operator statically pre-allocates an upper triangular all-ones matrix $\matU_s$ for all $s=32,64,128$. The baseline operator doesn't use the cube unit, while \MCScan~takes advantage of all the computing units reaching up to 37.5\% of theoretical memory bandwidth (peak bandwidth is 800GB/s for 910B4). To have a more solid evaluation of our implementation, we compare it to a \texttt{copy} kernel that performs a memory copy; we used the \texttt{torch.clone()}. A clear trend is that the larger the matrix multiplication dimension $s$ is, the better the performance of the multi-core scan. $s=128$ (almost) maximizes the utilization of the level-0 scratchpad memories $\LOA$ and $\LOB$ of the cube unit. We foresee that increasing the matrix multiplication tile size could lead to further performance improvements, but we leave this investigation to future work.

In addition, we investigate the additional performance benefits of taking advantage of the lower-precision input data (int8) capability of the cube unit; the interested reader is deferred to Section~\ref{appendix:mcscan_int8_fp16}. 
%
\subsubsection{Scan Improvements}\label{sec:mcscan:improve}
%
%
\begin{figure}[h]
\begin{center}
  \includegraphics[width=\columnwidth]{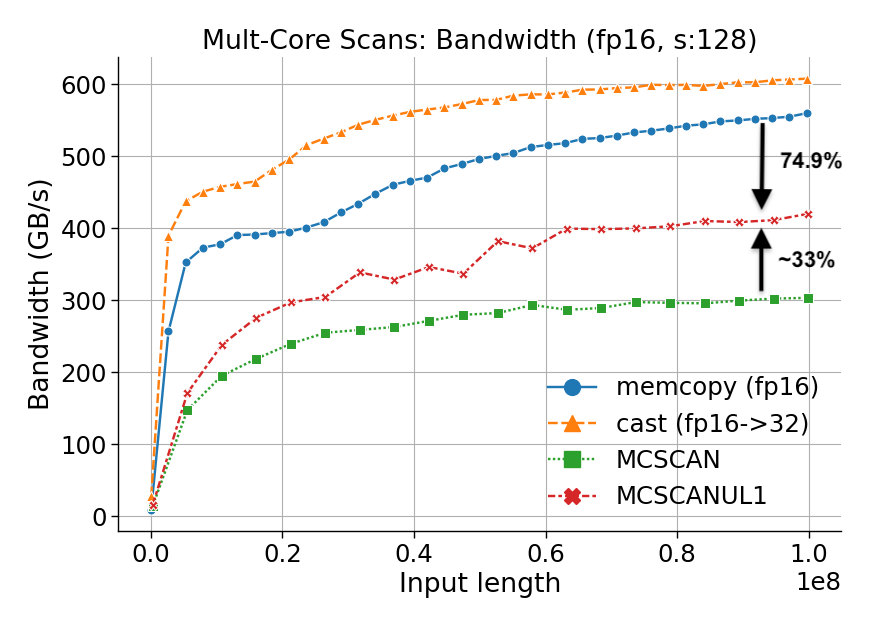}
\end{center}
\vspace{-7mm}
\caption{\small \MCScan~versus multi-core \ScanULONE}
\label{fig:exp:mcscan_versus_mcscanul1}
\end{figure}
\begin{figure}[h]
\begin{center}
  \includegraphics[width=\columnwidth]{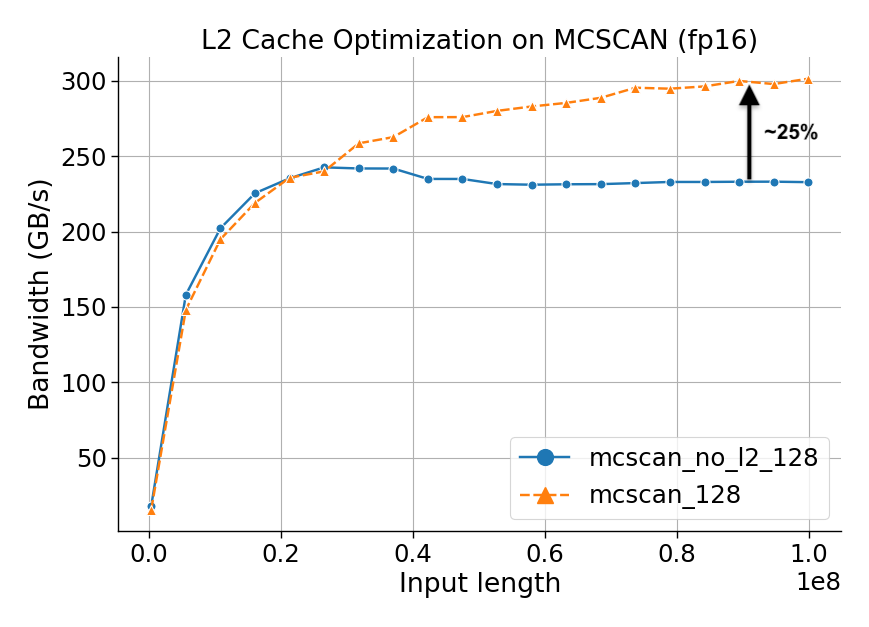}
\end{center}
\vspace{-8mm}
\caption{\small L2 cache optimization of \MCScan~($s=128$).}
\label{fig:exp:l2_cache_mcscan}
\end{figure}

%
We have implemented two optimizations for \MCScan. First, we implemented an L2 cache optimization that we call ``L2 cache splitting''. This L2 cache optimization splits the input into consecutive chunks so that the input and intermediate arrays fit in the L2 cache during the processing of each chunk. Each chunk is processed serially, and the last value of each chunk is propagated to the next chunk to carry ahead the partial prefix sums.~\Cref{fig:exp:l2_cache_mcscan} shows a performance improvement of 25\% for \MCScan~when the ``L2 cache splitting'' optimization is enabled. This optimization is already integrated in the \MCScan.
%
%

\begin{figure}[h]
\begin{center}
  \includegraphics[width=\columnwidth]{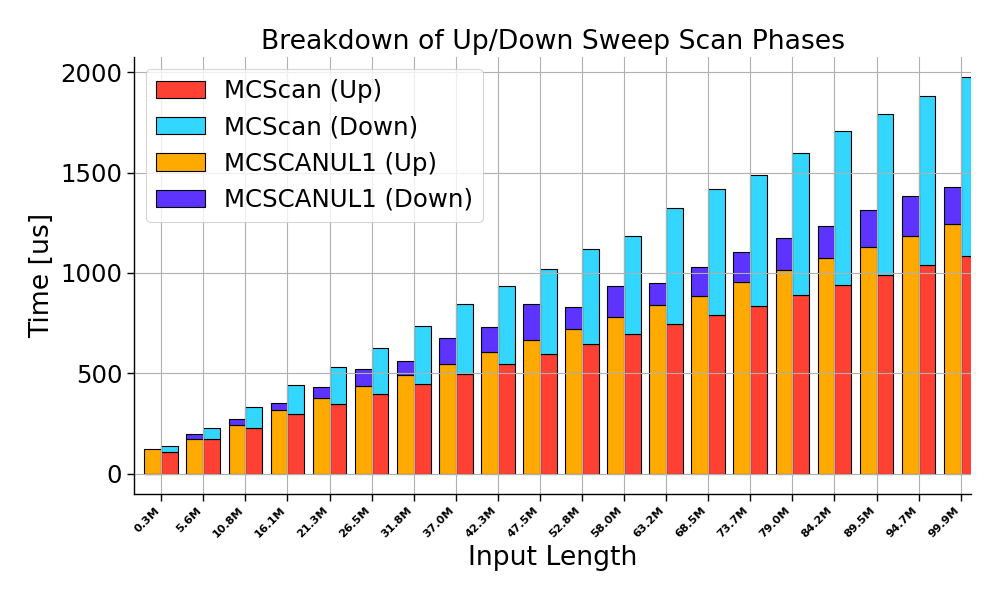}
\end{center}
\vspace{-8mm}
\caption{\small Break-down of up/down sweeps of scans.}
\label{fig:exp:breakdown_scans}
\end{figure}

Second, we designed and implemented a multi-core \ScanULONE~algorithm for fp16 by replacing the single matrix multiplication during Phase I of \MCScan~with the three matrix multiplications of \ScanULONE. We call the resulting algorithm, \MCScanULONE.~\Cref{fig:exp:mcscan_versus_mcscanul1} shows a performance improvement of 33\% of \MCScanULONE~versus \MCScan~and \MCScanULONE~reaches up to 74.9\% of the memcopy memory bandwidth as advertised in the abstract. The improvement comes from the fact that \ScanULONE~has an $\OO(s)$-depth speedup compared to \ScanU, see work/span discussion in Section~\ref{sec:single_cube_core_scan}.
In addition, we provide a performance breakdown analysis of the up-sweep (Phase I) and down-sweep (Phase II) phases of both \MCScan~and \MCScanULONE~in~\Cref{fig:exp:breakdown_scans}. The figure shows that, although the up-sweep phase of \MCScanULONE~is slightly longer than the up-sweep phase of \MCScan, the down-sweep phase of \MCScanULONE~is significantly shorter.
%
%
%
\subsection{Compress}
%
%
\begin{figure}[h]
\begin{center}
  \includegraphics[width=\vnlfigwidth]{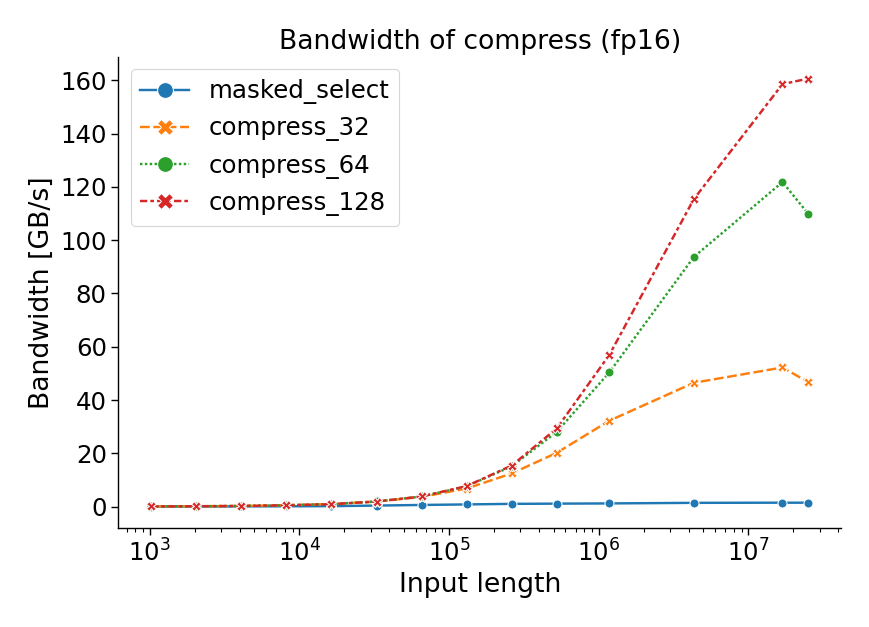}
\end{center}
\vspace{-7mm}
\caption{\small Bandwidth of compress operator based on \MCScan~($s=32,64,128$) against \texttt{torch.masked\_select}.}
\label{fig:exp:masked_select}
\end{figure}
%
%
\Cref{fig:exp:masked_select} depicts the performance comparison between compress versus the baseline PyTorch \texttt{masked\_select}. We set the mask so that each mask entry is independently set to true or false uniformly at random. The figure indicates that the baseline \texttt{masked\_select} operator is not optimized on Ascend, and a code investigation reveals that the baseline does not use the vector or cube units. On the other hand, our Compress kernel reaches up to 160GB/s (20\% of peak memory bandwidth).

%
\subsection{Radix Sort}
%
We modified the radix sort operator to additionally return indices that correspond to the input index of each output element. This modification ensures a fair comparison with the sort operator provided by the PyTorch Ascend adapter. The modification is based on the \texttt{SplitInd} operator by keeping track of the output indices on each split application. Our radix sort implementation is stable and supports unsigned (or signed) integers and floats (fp$16$).
%
\begin{figure}[ht]
\begin{center}
  \includegraphics[width=\columnwidth]{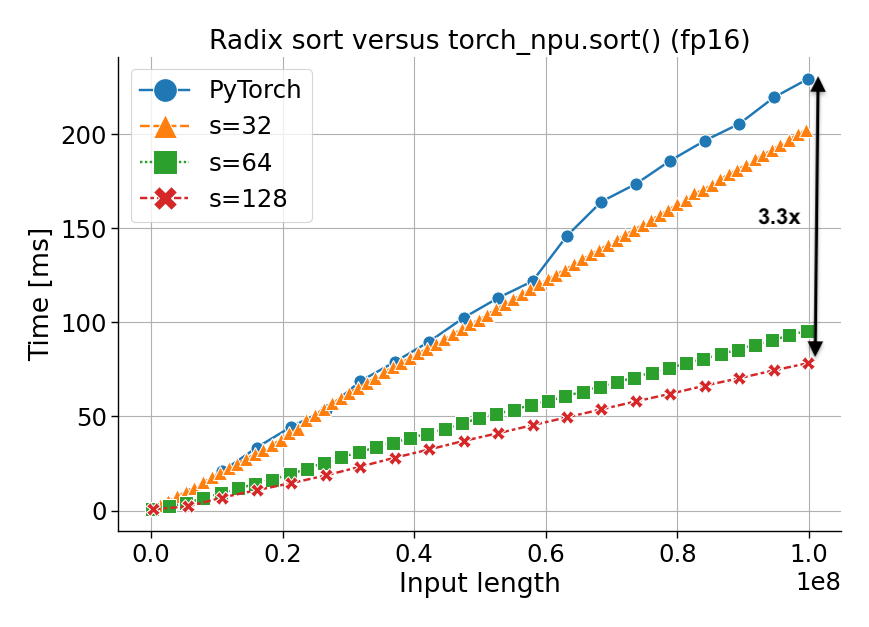}
\end{center}
\vspace{-7mm}
\caption{\small Comparison between radix sort and baseline \texttt{torch\_npu.sort()} for half-precision (fp$16$).}
\label{fig:exp:radixsort}
\end{figure}
%

%
\Cref{fig:exp:radixsort} shows the performance of a parallel fp16 radix sort implementation using \MCScan~with input data type int8 (\Cref{alg:scan_multi_core}) to perform the parallel split step. For input lengths greater than 525K, our ``textbook'' implementation of radix sort delivers a speedup between $1.3\times$ and $3.3\times$ compared to the \texttt{torch.sort()} baseline.
%

%
\paragraph{Performance Opportunities for small bit-width inputs}
%
We expect additional performance improvements for low-precision data types (low bit-width) since the number of radix sort iterations equals the input bit-width. Indeed, the trend in AI accelerators is to introduce low-precision formats like $8$-bit floats or $4$-bit integers (int4)~\cite{nvidiaA100}. Therefore, an additional performance improvement ($2\times$) for radix sorting in low-precision $8$-bit scenarios is expected without any extra development effort, see \Cref{fig:exp:radixsort_opportunity}.
%
\subsection{Batched Scan}
\begin{figure}[ht]
\begin{center}
  \includegraphics[width=\vnlfigwidth]{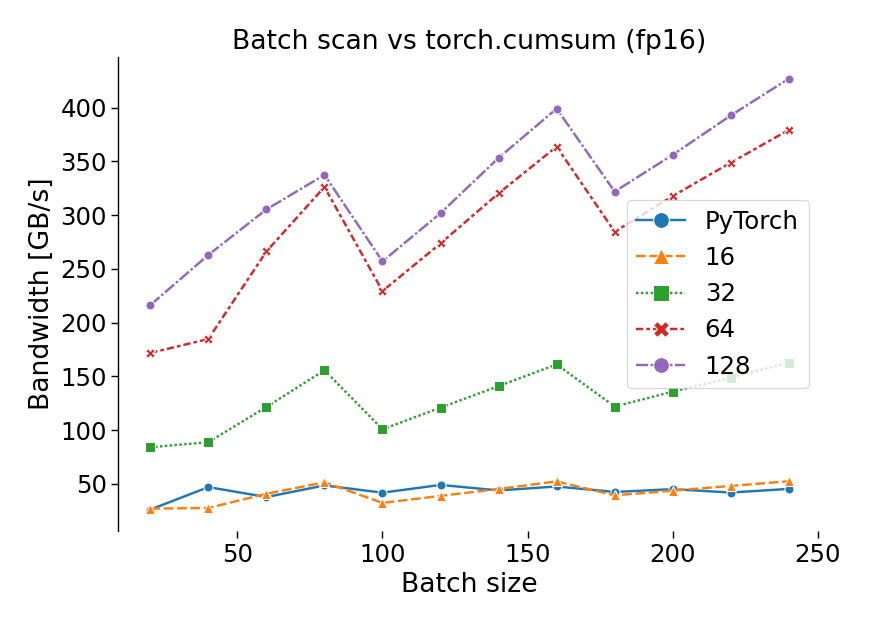}
\end{center}
\vspace{-7mm}
\caption{\small Bandwidth of batched scan based on~\Cref{alg:scan_single_core} for increasing batch sizes and $s=16,32,64,128$. Input length is $65K$.}
\label{fig:exp:batch_scan}
\end{figure}

%
\Cref{fig:exp:batch_scan} depicts the performance of the batched scan kernels on Ascend for varying input batch sizes and input lengths equal to $65K$. We depict the memory bandwidth achieved for tiling parameters $s=16,32,64$ and $128$. Our proposed batch scan operators for $s=64$ and $128$ reach up to $400$ GB/s. Interestingly enough, for smaller values of $s=16,32$, the performance of the proposed batch scan kernels is poor. In addition, the performance of our proposed batch scan kernel for $s=16$ and the baseline is similar. 
%
%
%
\subsection{Top-$p$ Sampling}
%
%
\begin{figure}[ht]
\begin{center}
\includegraphics[width=\vnlfigwidth]{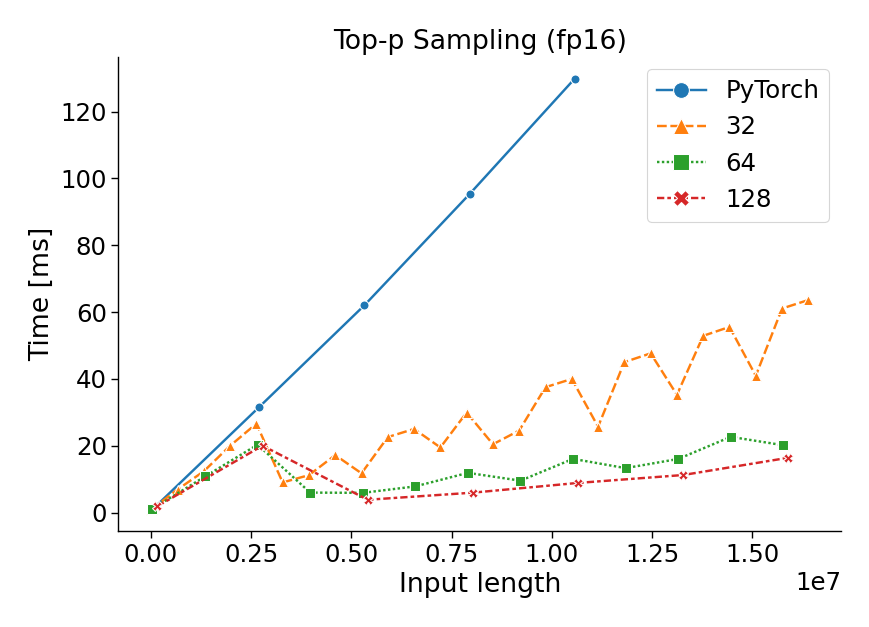}
\end{center}
\vspace{-7mm}
\caption{\small Execution time (in milliseconds) of top-$p$ sampling of Llama3 operator for a single batch. The baseline is labeled \texttt{PyTorch} and supports only discrete distributions with at most $2^{24}$ elements.}
\label{fig:exp:top_p}
\end{figure}
%
\Cref{fig:exp:top_p} depicts the execution time of drawing one sample using the top-$p$ sampler as it is implemented in the Llama3 model, see~\cite{llama_generation_2024}. PyTorch corresponds to the Ascend implementation using the baseline sort and cumsum operators. The lineplots with labels $s=32,64,128$ are similar to the baseline by replacing the \texttt{sort()} and \texttt{cumsum()} operators with the proposed radix sort and multi-core scan, respectively. The figure shows that the baseline top-$p$ sampling implementation scales poorly, mainly because the baseline \texttt{torch.cumsum} operator is not optimized for Ascend.
%
%
%
\section{Conclusion}\label{sec:conclusion}
%
We developed and evaluated efficient parallel scan algorithms tailored to the Ascend architecture by leveraging the power of the cube (matrix multiplication) unit. Our results demonstrate substantial performance improvements, with speedups ranging from $5\times$ to $9.6\times$ compared to vector-only implementations for sufficiently large input lengths. Additionally, we presented a multi-core Ascend scan algorithm that fully utilizes both the cube and vector units of Ascend, reaching up to 74.9\% of the theoretical memory bandwidth.
Last, we extended our contributions to include crucial computational kernels for AI workloads, such as parallel split, compress/compact, and top-$p$ (nucleus) sampling, all exhibiting significant performance gains. Furthermore, our optimized implementation of radix sort, which utilizes matrix multiplications for parallel splits, showcases the potential of matrix engines in enhancing complex operations, offering up to $3.3\times$ speedup over the baseline.
%
\clearpage
{\small
\bibliographystyle{siamplain}

\begin{thebibliography}{10}

\bibitem{amd_matrix_cores}
{\sc {AMD}}, {\em {AMD} {CDNA} 2 {A}rchitecture}, technical report, Advanced Micro Devices, Inc., 2022, \url{https://www.amd.com/system/files/documents/amd-cdna2-white-paper.pdf}.
\newblock Accessed: 2024-09-09.

\bibitem{model:PEM:arge08}
{\sc L.~Arge, M.~T. Goodrich, M.~Nelson, and N.~Sitchinava}, {\em Fundamental parallel algorithms for private-cache chip multiprocessors}, in Proceedings of Symposium on Parallelism in Algorithms and Architectures (SPAA), SPAA '08, New York, NY, USA, 2008, ACM, p.~197–206, \url{https://doi.org/10.1145/1378533.1378573}.

\bibitem{armv9:manual_2024}
{\sc ARM}, {\em {ARM} architecture reference manual for a-profile architecture}, 2024, \url{https://developer.arm.com/documentation/ddi0487/ka/?lang=en}.

\bibitem{scan:moderngpu}
{\sc S.~Baxter}, {\em moderngpu 2.0}.
\newblock \url{https://github.com/moderngpu/moderngpu/wiki}, 2016.

\bibitem{cuda:thrust}
{\sc N.~Bell and J.~Hoberock}, {\em Chapter 26 - thrust: A productivity-oriented library for cuda}, in GPU Computing Gems Jade Edition, W.~mei W.~Hwu, ed., Applications of GPU Computing Series, Morgan Kaufmann Publishers Inc., Boston, 2012, pp.~359--371, \url{https://doi.org/10.1016/B978-0-12-385963-1.00026-5}.

\bibitem{scan:blelloch_iee89}
{\sc G.~E. Blelloch}, {\em Scans as primitive parallel operations}, IEEE Transactions on Computers, 38 (1989), pp.~1526--1538, \url{https://doi.org/10.1109/12.42122}.

\bibitem{blelloch_prefix_1990}
{\sc G.~E. Blelloch}, {\em Prefix sums and their applications}, in Sythesis of parallel algorithms, Morgan Kaufmann Publishers Inc., San Francisco, CA, USA, 1st~ed., 1990, pp.~35--60.

\bibitem{book:blelloch_vector_models}
{\sc G.~E. Blelloch}, {\em Vector models for data-parallel computing}, MIT Press, Cambridge, MA, USA, 1990.

\bibitem{prog:blelloch_cacm96}
{\sc G.~E. Blelloch}, {\em Programming parallel algorithms}, Commun. ACM, 39 (1996), p.~85–97, \url{https://doi.org/10.1145/227234.227246}.

\bibitem{radix_sort:spaa91_CM2}
{\sc G.~E. Blelloch, C.~E. Leiserson, B.~M. Maggs, C.~G. Plaxton, S.~J. Smith, and M.~Zagha}, {\em A comparison of sorting algorithms for the connection machine cm-2}, in Proceedings of Symposium on Parallelism in Algorithms and Architectures (SPAA), SPAA '91, New York, NY, USA, 1991, Association for Computing Machinery, p.~3–16, \url{https://doi.org/10.1145/113379.113380}.

\bibitem{tcu_model:spaa20}
{\sc R.~Chowdhury, F.~Silvestri, and F.~Vella}, {\em A computational model for tensor core units}, in Proceedings of Symposium on Parallelism in Algorithms and Architectures (SPAA), Philadelphia, USA, 2020, ACM, p.~519–521, \url{https://doi.org/10.1145/3350755.3400252}.

\bibitem{tcu_model:europar21}
{\sc R.~Chowdhury, F.~Silvestri, and F.~Vella}, {\em Algorithm design for tensor units}, in International Conference on Parallel and Distributed Computing (Euro-Par), Lisbon, Portugal, 2021, Springer-Verlag, p.~353–367.

\bibitem{scan:tcuICS19}
{\sc A.~Dakkak, C.~Li, J.~Xiong, I.~Gelado, and W.-m. Hwu}, {\em Accelerating reduction and scan using tensor core units}, in Proceedings of the ACM International Conference on Supercomputing (ICS), ICS '19, Phoenix AZ, 2019, ACM, p.~46–57, \url{https://doi.org/10.1145/3330345.3331057}.

\bibitem{matrix_engines_debate}
{\sc J.~Domke, E.~Vatai, A.~Drozd, P.~ChenT, Y.~Oyama, L.~Zhang, S.~Salaria, D.~Mukunoki, A.~Podobas, M.~WahibT, and S.~Matsuoka}, {\em Matrix engines for high performance computing: A paragon of performance or grasping at straws?}, in International Conference on Parallel \& Distributed Processing Symposium (IPDPS), Los Alamitos, CA, USA, May 2021, IEEE Computer Society, pp.~1056--1065, \url{https://doi.org/10.1109/IPDPS49936.2021.00114}.

\bibitem{scan:matrixscan08}
{\sc Y.~Dotsenko, N.~K. Govindaraju, P.-P. Sloan, C.~Boyd, and J.~Manferdelli}, {\em Fast scan algorithms on graphics processors}, in Proceedings of the ACM International Conference on Supercomputing (ICS), ICS '08, New York, NY, USA, 2008, ACM, p.~205–213, \url{https://doi.org/10.1145/1375527.1375559}.

\bibitem{gu2024mamba}
{\sc A.~Gu and T.~Dao}, {\em Mamba: Linear-time sequence modeling with selective state spaces}, 2024, \url{https://openreview.net/forum?id=AL1fq05o7H}.

\bibitem{NaN2024ScanSpeed}
{\sc J.~Hemstad}, {\em Lecture 24: Scan at the speed of light}.
\newblock {YouTube video}, July 2024, \url{https://www.youtube.com/watch?v=VLdm3bV4bKo}.
\newblock Online; posted on YouTube.

\bibitem{top_p_sampling}
{\sc A.~Holtzman, J.~Buys, L.~Du, M.~Forbes, and Y.~Choi}, {\em The curious case of neural text degeneration}, in International Conference on Learning Representations (ICLR), Addis Ababa, Ethiopia, 2020, OpenReview.net, \url{https://openreview.net/forum?id=rygGQyrFvH}.

\bibitem{scan:par_weight_sampling}
{\sc L.~H\"{u}bschle-Schneider and P.~Sanders}, {\em Parallel weighted random sampling}, ACM Trans. Math. Softw., 48 (2022), \url{https://doi.org/10.1145/3549934}.

\bibitem{book:gpus:scan_chapter}
{\sc W.~W. Hwu, D.~B. Kirk, and I.~{El Hajj}}, {\em Chapter 11 - prefix sum (scan)}, in Programming Massively Parallel Processors (Fourth Edition), Morgan Kaufmann Publishers Inc., Cambridge, Massachusetts, United States, fourth~ed., 2023, pp.~253--256, \url{https://doi.org/10.1016/B978-0-323-91231-0.00006-9}.

\bibitem{book:gpus}
{\sc W.~W. Hwu, D.~B. Kirk, and I.~{El Hajj}}, {\em Programming Massively Parallel Processors (Fourth Edition)}, Morgan Kaufmann Publishers Inc., fourth~ed., 2023, \url{https://doi.org/10.1016/B978-0-323-91231-0.00006-9}.

\bibitem{intel_isa_extensions}
{\sc {Intel Corporation}}, {\em Intel architecture instruction set extensions programming reference}, technical report, Intel Corporation, 2022.
\newblock Accessed: 2024-09-09.

\bibitem{topK:database21}
{\sc J.~Johnson, M.~Douze, and H.~Jégou}, {\em Billion-scale similarity search with {GPUs}}, IEEE Transactions on Big Data, 7 (2021), pp.~535--547, \url{https://doi.org/10.1109/TBDATA.2019.2921572}.

\bibitem{google_tpus:ISCA17}
{\sc N.~P. Jouppi and et~al.}, {\em In-datacenter performance analysis of a tensor processing unit}, in Proceedings of International Symposium on Computer Architecture (ISCA), Toronto, ON, Canada, 2017, ACM, p.~1–12, \url{https://doi.org/10.1145/3079856.3080246}.

\bibitem{tpu:v4}
{\sc N.~P. Jouppi and et~al.}, {\em Ten lessons from three generations shaped google’s {TPUv4i}: Industrial product}, in Proceedings of International Symposium on Computer Architecture (ISCA), Online -- Worldwide, 2021, {IEEE}, pp.~1--14, \url{https://doi.org/10.1109/ISCA52012.2021.00010}.

\bibitem{book:knuth:artv3_sorting}
{\sc D.~E. Knuth}, {\em The art of computer programming: sorting and searching}, vol.~3, Addison-Wesley Publishing Co., USA, 2nd~ed., 1998.

\bibitem{gpu_model_ipdps14}
{\sc A.~Koike and K.~Sadakane}, {\em A novel computational model for {GPUs} with application to {I/O} optimal sorting algorithms}, in International Conference on Parallel \& Distributed Processing Symposium (IPDPS), Milan, Italy, 2014, IEEE, pp.~614--623, \url{https://doi.org/10.1109/IPDPSW.2014.72}.

\bibitem{topK:ICML19}
{\sc W.~Kool, H.~Van~Hoof, and M.~Welling}, {\em Stochastic beams and where to find them: The {G}umbel-top-k trick for sampling sequences without replacement}, in International Conference on Machine Learning (ICML), K.~Chaudhuri and R.~Salakhutdinov, eds., vol.~97 of Proceedings of Machine Learning Research, Long Beach, California, USA, June 2019, PMLR, pp.~3499--3508, \url{https://proceedings.mlr.press/v97/kool19a.html}.

\bibitem{kung1978systolic}
{\sc H.~Kung and C.~Leiserson}, {\em Systolic Arrays for ({VLSI})}, CMU-CS, Carnegie-Mellon University, Department of Computer Science, 1978.

\bibitem{book:scan94}
{\sc S.~Lakshmivarahan and S.~K. Dhall}, {\em Parallel Computing Using the Prefix Problem}, Oxford University Press, Oxford, United Kingdom, 1994.

\bibitem{topK:poster_PPoPP24}
{\sc Y.~Li, B.~Zhou, J.~Zhang, X.~Wei, Y.~Li, and Y.~Chen}, {\em Poster: {RadiK}: Scalable radix top-k selection on {GPUs}}, in Proceedings of the ACM Symposium on Principles and Practice of Parallel Programming (PPoPP), PPoPP '24, New York, NY, USA, 2024, ACM, p.~472–474, \url{https://doi.org/10.1145/3627535.3638478}.

\bibitem{topK:radik_ICS24}
{\sc Y.~Li, B.~Zhou, J.~Zhang, X.~Wei, Y.~Li, and Y.~Chen}, {\em {RadiK}: Scalable and optimized {GPU}-parallel radix top-{K} selection}, in Proceedings of the ACM International Conference on Supercomputing (ICS), ICS '24, New York, NY, USA, 2024, ACM, p.~537–548, \url{https://doi.org/10.1145/3650200.3656596}.

\bibitem{book:ascend_cann_xiaoyao20}
{\sc X.~Liang}, {\em Ascend AI Processor Architecture and Programming: Principles and Applications of CANN}, Elsevier Science, 2020.

\bibitem{huawei_ascend:hpca2021}
{\sc H.~Liao, J.~Tu, J.~Xia, H.~Liu, X.~Zhou, H.~Yuan, and Y.~Hu}, {\em Ascend: a scalable and unified architecture for ubiquitous deep neural network computing: industry track paper}, in Proceedings of International Symposium on High-Performance Computer Architecture (HPCA), Seoul, South Korea, 2021, {IEEE}, pp.~789--801, \url{https://doi.org/10.1109/HPCA51647.2021.00071}.

\bibitem{huawei_ascend:hotchips19}
{\sc H.~Liao, J.~Tu, J.~Xia, and X.~Zhou}, {\em {DaVinci}: A scalable architecture for neural network computing}, in Hot Chips: A Symposium on High-Performance Chips (HCS), Cupertino, CA, USA, 2019, {IEEE}, pp.~1--44, \url{https://doi.org/10.1109/HOTCHIPS.2019.8875654}.

\bibitem{quant_llm:mlsys24_lin2023awq}
{\sc J.~Lin, J.~Tang, H.~Tang, S.~Yang, W.-M. Chen, W.-C. Wang, G.~Xiao, X.~Dang, C.~Gan, and S.~Han}, {\em {AWQ}: Activation-aware weight quantization for {LLM} compression and acceleration}, in MLSys, Santa Clara, California, USA, 2024, mlsys.org, pp.~1--14.

\bibitem{search_pram_matias97}
{\sc Y.~Matias}, {\em Parallel algorithms column: on the search for suitable models}, SIGACT News, 28 (1997), p.~21–29, \url{https://doi.org/10.1145/262301.262305}.

\bibitem{scan:decouple_lookback}
{\sc D.~Merrill and M.~Garland}, {\em Single-pass parallel prefix scan with decoupled lookback}, in Not available, Santa Clara, CA, USA, 2016, NVIDIA, pp.~1--9, \url{https://research.nvidia.com/publication/2016-03_single-pass-parallel-prefix-scan-decoupled-look-back}.

\bibitem{llama_generation_2024}
{\sc {{Meta AI}}}, {\em Llama3 generation code - sample\_top\_p() method}.
\newblock \url{https://github.com/meta-llama/llama3/blob/main/llama/generation.py\#L358}, 2024.
\newblock Accessed: 2024-09-24.

\bibitem{nvidia:cub}
{\sc {NVIDIA}}, {\em Cooperative primitives for {CUDA} {C++}}, 2023, \url{https://github.com/NVIDIA/cub}.

\bibitem{nvidia:tesla_v100}
{\sc {NVIDIA Authors}}, {\em {NVIDIA} {DGX-1} with {T}esla {V100} system architecture}, Tech. Report MSU-CSE-06-2, Nvidia Corporation, Dec. 2017, \url{https://images.nvidia.com/content/pdf/dgx1-v100-system-architecture-whitepaper.pdf}.

\bibitem{nvidiaA100}
{\sc {NVIDIA Corporation}}, {\em {NVIDIA} {A100} Tensor Core {GPU} Architecture}, 2020.
\newblock \url{https://www.nvidia.com/content/dam/en-zz/Solutions/Data-Center/nvidia-ampere-architecture-whitepaper.pdf}.

\bibitem{gpu_gems_book_2}
{\sc M.~Pharr and R.~Fernando}, {\em {GPU Gems} 2: Programming Techniques for High-Performance Graphics and General-Purpose Computation ({GPU Gems})}, Addison-Wesley Publishing Co., Boston, USA, 2005.

\bibitem{scan:gpu_harris07}
{\sc S.~Sengupta, M.~Harris, Y.~Zhang, and J.~D. Owens}, {\em Scan primitives for {GPU} computing}, in Proceedings of the Symposium on Graphics Hardware (EuroGraphics), GH '07, Goslar, DEU, 2007, Eurographics Association, p.~97–106.

\bibitem{provably_gpu_algos}
{\sc N.~Sitchinava and V.~Weichert}, {\em Provably efficient {GPU} algorithms}, CoRR, abs/1306.5076 (2013), pp.~1--25, \url{https://arxiv.org/abs/1306.5076}.

\bibitem{scan:snir86}
{\sc M.~Snir}, {\em Depth-size trade-offs for parallel prefix computation}, Journal of Algorithms, 7 (1986), pp.~185--201, \url{https://doi.org/10.1016/0196-6774(86)90003-9}.

\bibitem{algo_parallel_computers1985}
{\sc L.~Snyder, L.~H. Jamieson, D.~B. Gannon, and H.~J. Siegel}, {\em Algorithmically Specialized Parallel Computers}, Academic Press, Cambridge, Massachusetts, United States, 1985.

\bibitem{ibm_power10_mma}
{\sc W.~J. Starke, B.~W. Thompto, J.~A. Stuecheli, and J.~E. Moreira}, {\em {IBM}'s {POWER10} processor}, IEEE Micro, 41 (2021), pp.~7--14, \url{https://doi.org/10.1109/MM.2021.3058632}.

\bibitem{llama}
{\sc H.~Touvron, T.~Lavril, G.~Izacard, X.~Martinet, M.-A. Lachaux, T.~Lacroix, B.~Rozière, N.~Goyal, E.~Hambro, F.~Azhar, A.~Rodriguez, A.~Joulin, E.~Grave, and G.~Lample}, {\em Llama: Open and efficient foundation language models}, 2023, \url{https://arxiv.org/abs/2302.13971}.

\bibitem{streamscan_ppopp13}
{\sc S.~Yan, G.~Long, and Y.~Zhang}, {\em {StreamScan}: fast scan algorithms for gpus without global barrier synchronization}, SIGPLAN Not., 48 (2013), p.~229–238, \url{https://doi.org/10.1145/2517327.2442539}.

\bibitem{topK:survey_SC23}
{\sc J.~Zhang, A.~Naruse, X.~Li, and Y.~Wang}, {\em Parallel top-k algorithms on {GPU}: A comprehensive study and new methods}, in Proceedings of the International Conference for High Performance Computing, Networking, Storage and Analysis (SC), SC '23, New York, NY, USA, 2023, ACM, \url{https://doi.org/10.1145/3581784.3607062}.

\bibitem{scan:zero_def_circuits06}
{\sc H.~Zhu, C.-K. Cheng, and R.~Graham}, {\em On the construction of zero-deficiency parallel prefix circuits with minimum depth}, ACM Trans. Des. Autom. Electron. Syst., 11 (2006), p.~387–409, \url{https://doi.org/10.1145/1142155.1142162}.

\bibitem{scan:zouzias_europar23}
{\sc A.~Zouzias and W.~F. McColl}, {\em A parallel scan algorithm in the tensor core unit model}, in International Conference on Parallel and Distributed Computing (Euro-Par), vol.~14100 of Lecture Notes in Computer Science, Limassol, Cyprus, 2023, Springer, pp.~489--502, \url{https://doi.org/10.1007/978-3-031-39698-4\_33}.

\end{thebibliography}

}
%
%

\appendix

%
\section{AscendC Programming Model}\label{sec:ascendc}
%
Recently, a pipeline-based programming model for Ascend called \emph{AscendC} has been developed. AscendC allows its users to build high-performance computational kernels for the Ascend architecture. Such kernels are usually called AscendC operators. The AscendC programming model is built on top of C++, and it allows its users to have fine-grained control of Ascend's hardware components, such as MTEs, scalar, vector, and cube compute engines. At the same time, AscendC eliminates many potential problems, such as the need to explicitly synchronize hardware components within AIC and AIV cores.

The AscendC programming model is based on a multiple pipeline abstraction model. AscendC provides users with some abstractions, including a context manager object, tensors, queues, and buffers.

AscendC provides tensor structures as wrappers over data allocated in the global or core's local memory. \emph{GlobalTensor} is a structure that represents a buffer in global memory. In all operators, both input and output data come from global tensors. On the other hand, \emph{LocalTensor} represents a buffer in the core's local memory. Users can allocate local tensors in one of multiple possible hardware buffers (UB, L1, L0A, L0B, etc.).

AscendC also provides the \emph{queues} API -- data structures used for managing tensors and resolving data dependencies between different hardware components which work on the same tensors. After a hardware component interacts with a tensor the \texttt{Enque} method is called, that saves the pointer into the queue. Then when the next hardware component needs to interact with the same tensor the \texttt{Deque} method is called, that waits for the corresponding \texttt{Enque} to be called, ensuring synchronization, and returns the pointer to the local tensor. This way, all data dependencies are explicit, and the computational pattern is consistent across all local tensors and all physical buffers. Queues can contain more than one tensor at a time -- in many cases, implementing double buffering comes down to changing the queue capacity from the default value one to two.
Naturally, the model also defines dozens of possible operations on tensors; we only mention some of the most frequently used here:
\begin{itemize}
    \item \emph{DataCopy}. MTE's function that copies data from an input to an output tensor. The basic version copies a number of continuous elements but can also be configured for strides and automatic layout transformations.
    \item \emph{Mmad}. AIC core's function that multiplies two input matrices (local tensors) and writes the result to the output matrix (a local tensor). The result can be accumulated with existing values in the output tensor.
    \item \emph{Adds}. AIV core's function that adds a scalar to an input local tensor and writes the result to the output local tensor.
    \item \emph{GatherMask}. AIV core's function that takes an input local tensor and a binary mask, also a local tensor, and gathers all the elements from the input tensor for which the corresponding value in the mask is equal to $1$. Gathered elements are stored in the contiguous form in the output local tensor.
\end{itemize}

An operator is executed using multiple \emph{blocks} -- block is the smallest logical execution unit. The user specifies the number of blocks to be used when running the kernel. 
Another critical function AscendC provides is hardware synchronization among computing units -- \emph{SyncAll} allows the user to synchronize all blocks. The execution is continued only after each unit reaches the synchronization point. 

%
\section{Additional Material for Scans}\label{appendix:additional_material}
%
%
%
%
\subsection{Multiple (Batched) Scans}\label{sec:batched_scan}
%
The batched scan computes the prefix sum of a batch of input arrays of equal length in parallel. Given a scan algorithm for a 1D array, a corresponding batched scan algorithm could be defined by deciding how to schedule the cube and vector computations into the multiple cores.
%
\begin{figure}[h!]
\begin{center}
  \includegraphics[width=\columnwidth]{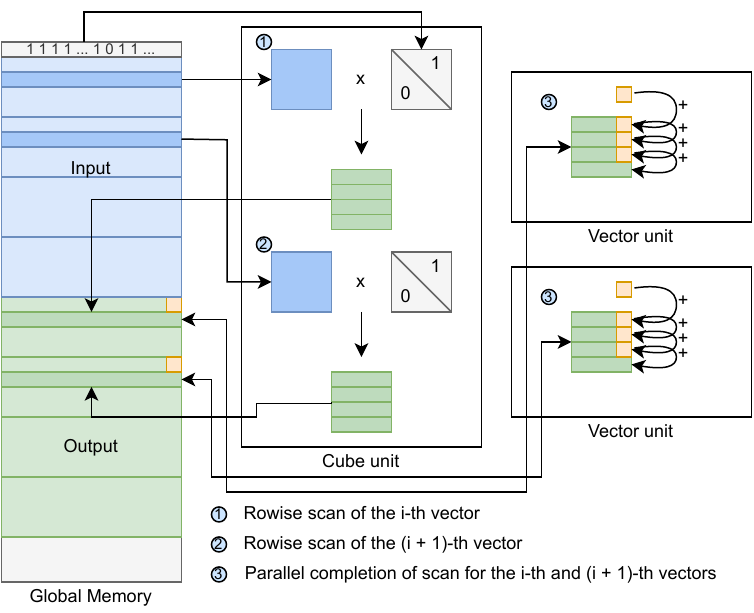}
\end{center}
\vspace{-7mm}
\caption{\small Batched scan algorithm based on \ScanU. Blue and green denote the input and output array, respectively.}
\label{fig:scan_batch}
\end{figure}
%

Let's discuss, as an example, a particular case where \ScanU~is used as a building block. The batched algorithm uses the same principles as the~\Cref{alg:scan_single_core} but also considers the 2-to-1 ratio between the vector and cube cores in the split Ascend architecture (910B).~\Cref{fig:scan_batch} depicts the main ideas behind our batched scan algorithm in this case. The algorithm starts by computing the local scans of size $s$ of $\x$ of all input arrays. Each cube core computes the local scans of a tile of size $\ell$ in two batches at the same time; once the tiles of the first two batches are ready, two distinct vector cores will complete the scans independently over each batch by propagating the partial sums within the tiles. This process is pipelined through AscendC to efficiently use all available hardware (vector) resources. The second batched scan algorithm extends \ScanULONE~(\Cref{alg:scan_cube_only}) so that each AI core computes a scan on a separate array in the batch.
%
\begin{figure}[ht]
\begin{center}
  \includegraphics[width=\columnwidth]{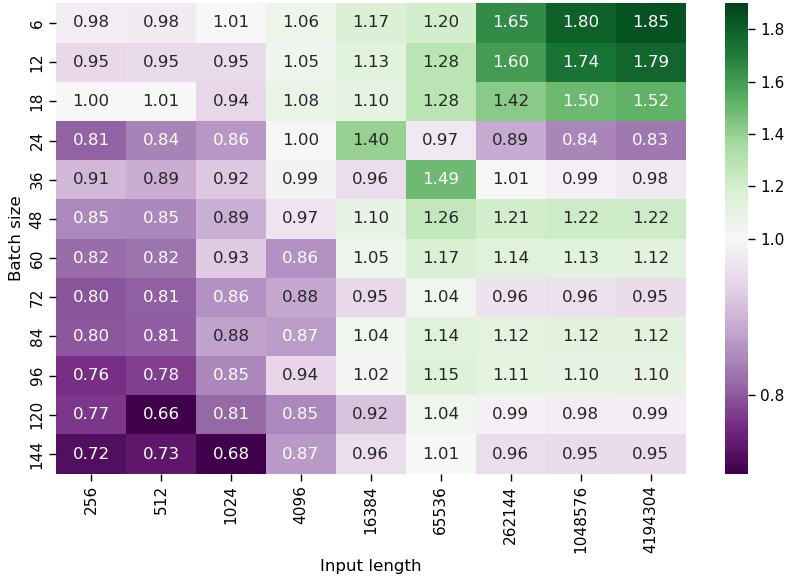}
\end{center}
\vspace{-7mm}
\caption{\small Execution time ratio between \ScanULONE~and \ScanU~batched scan algorithms for various array length ($x$-axis) and batch sizes ($y$-axis). Baseline is \ScanU.}
\label{fig:batch_scan_compare}
\end{figure}
%

\Cref{fig:batch_scan_compare} compares the two batched scan algorithms presented for increasing input batch size and array length. The first batched scan algorithm, based on \ScanU~(\Cref{alg:scan_single_core}), is used as our reference/baseline. Both algorithms have the same tiling strategy based on the input shapes to ensure a fair comparison. The figure demonstrates that the algorithms perform well in different cases and, more importantly, complement each other. In particular, \ScanU~is superior when the batch size is greater than $18$, and the input length is smaller than $4K$. On the other hand, \ScanULONE~is superior when the batch size is smaller than $18$ and the input length larger than $4K$. %
\subsection{Exclusive and int8 scan}\label{appendix:exclusive_mcscan_int8}
%
Here, we discuss a few extensions of the multi-core scan algorithm that we have implemented. In particular, we added support for exclusive scans and integer inputs. Typically, AI accelerators support low-precision arithmetic since inference of deep learning models is robust to extreme levels of quantization~\cite{quant_llm:mlsys24_lin2023awq}. In particular, Ascend supports input matrices of 8-bit integers with output/accumulation in 32-bit integers. Since scan is a memory-bound operator, there is an opportunity to improve performance in terms of elements per second processed; see~\Cref{fig:exp:mc_scan_fp16_vs_int8}. We have implemented a specialization of~\Cref{alg:scan_multi_core} for integers with 8 bits, which is extensively used in our scan applications.
We implemented exclusive scan by writing the output of the inclusive scan to global memory shifted by one element, discarding the last value and writing zero to the first position using the first block.
%
\subsection{Weighted Sampling}\label{sec:weighted_sampling}
%
We implement a parallel weighted sampling kernel using the well-known inverse transform sampling approach and a scan to compute the cumulative distribution. Given an array $\w$ of $n$ positive weights, the goal is to draw a sample with proportional probability to the weights. The output is an index $i$ of $\w$ with probability proportional to $w_i$. First, we scan $\w$, and then, given a uniform sample $\theta\in{[0,1]}$, we invoke the \texttt{SplitInd} kernel with input $\text{scan}(\w)$ and the element-wise predicate $ ? > \theta * \sum_{i}{w_i}$. The last entry of the output indices array of \texttt{SplitInd} contains the weighted sample. For more advanced parallel weighted sampling techniques, see~\cite{scan:par_weight_sampling}.
The performance improvement of our proposed weighted sampling kernel is not significant compared to the baseline for a single sample. However, our implementation does provide a functional improvement compared to the baseline operator. Indeed, the baseline Ascend weighted sampling operator \texttt{torch.multinomial} supports discrete distributions with support size up to $2^{24}$ elements, whereas our approach can support distributions with arbitrary support size. We leave as future work any further possible improvements on parallel weighted sampling. In particular, for the multiple sample generation scenario, the parallel alias table construction of~\cite{scan:par_weight_sampling} seems to be a promising direction.
%
\subsection{Scan on low bit-width data types}\label{appendix:mcscan_int8_fp16}
\begin{figure}[ht]
\begin{center}
  \includegraphics[width=\columnwidth]{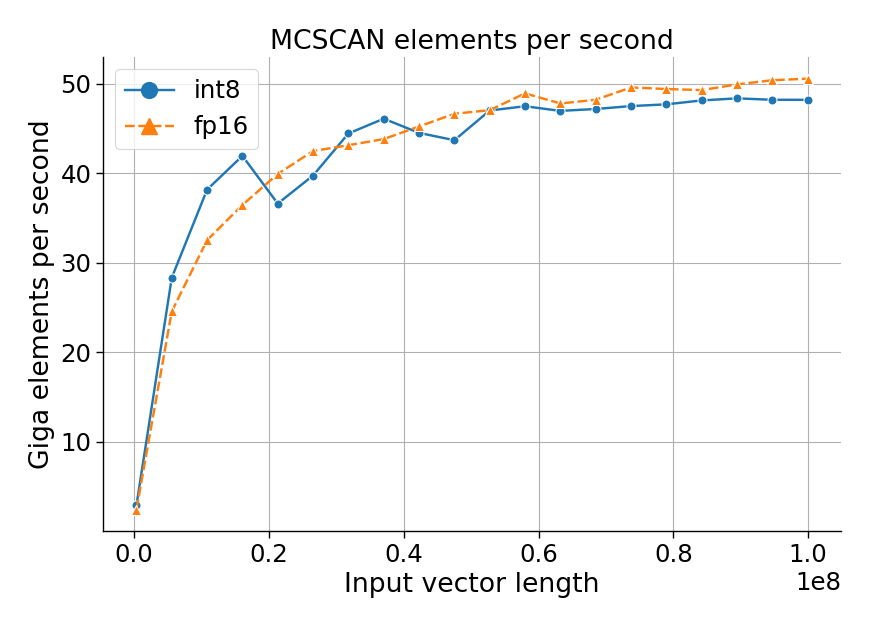}
\end{center}
\vspace{-8mm}
\caption{\small Giga elements per second comparison of \MCScan~for float16 (fp16) and 8-bit integers (int8) input data types.}
\label{fig:exp:mc_scan_fp16_vs_int8}
\end{figure}

%
%
Next, we investigate the additional performance benefits of taking advantage of the lower-precision input data (int8) capability of the cube unit.~\Cref{fig:exp:mc_scan_fp16_vs_int8} depicts the performance of the multi-core scan algorithm in terms of giga elements per second for input data types float16 and $8$-bit integers (int8). As depicted, there is only a small performance improvement for relatively small length, which is mainly due to our current implementation. For example, additional performance can be achieved since the Cube unit can multiply, in a single instruction, matrices of sizes $128\times 256$ and $256 \times 128$ in int8. 
%
%
%
\subsection{Radix sort on low bit-width inputs}\label{appendix:radix_sort_opp}
\begin{figure}[h]
\begin{center}
  \includegraphics[width=\columnwidth]{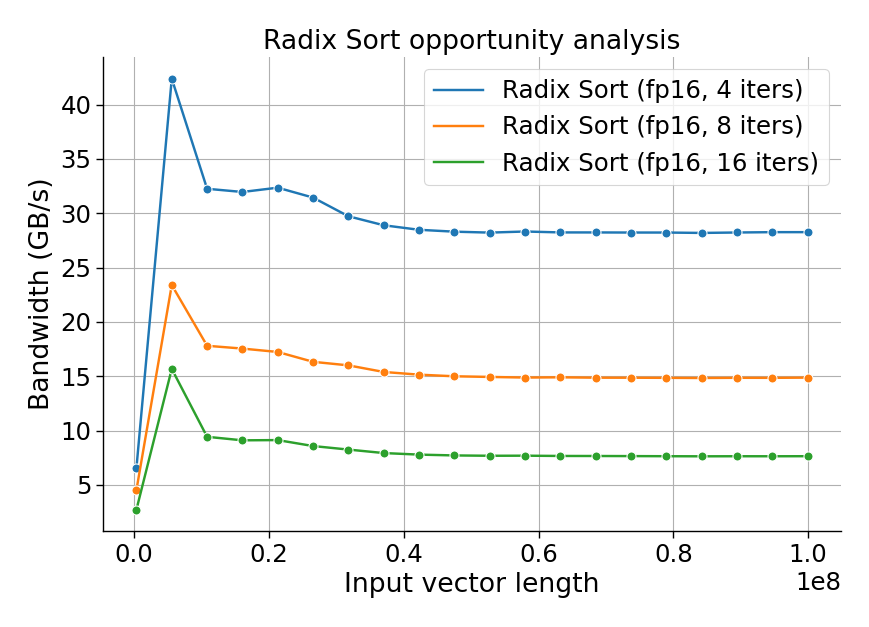}
\end{center}
\vspace{-7mm}
\caption{\small Radix sort opportunity analysis for low bit-width inputs.}
\label{fig:exp:radixsort_opportunity}
\end{figure}

%
%
Here, we provide an estimate of the expected performance improvements of radix sort for the case where the input data type has $4$ or $8$ bit-width. To do so, we measure the execution time of our radix sort kernel by reducing the number of radix sort iterations from $16$ to $4$ or $8$ (algorithmic correctness is not guaranteed). Notice that we still read and write using 16 bits per element.~\Cref{fig:exp:radixsort_opportunity} shows a significant performance improvement of roughly $1.5\times$ and $3\times$ of our radix sort kernel for $8$ and $4$ bits inputs, respectively.
%
%

\end{document}